\documentclass[11pt,a4paper]{article}
\pdfoutput=1
\usepackage[pdftex]{graphics}
\usepackage{jheppub}
\usepackage{amsfonts,amsmath,amssymb}
\usepackage{mathrsfs}
\usepackage{graphicx}
\graphicspath{{./Figures/}}
\usepackage{slashed}
\usepackage{booktabs}
\usepackage{dsfont}
\usepackage{mathtools}
\usepackage{bm}
\usepackage{tikz}
\usepackage{arydshln}   
\usepackage{wasysym}
\usepackage{tikz}
\usetikzlibrary{matrix}
\usetikzlibrary{shapes.arrows,chains}
\usepackage{subcaption}
\usepackage{multirow}
\usepackage{enumitem}
\usepackage{epsdice}

\newcommand{\nn}{\nonumber}

\def\half{\frac{1}{2}}

\def\det{{\rm det}}

\def\CA{{\cal A}}

\def\CF{{\cal F}}

\def\CH{{\cal H}}

\def\CL{{\cal L}}

\def\CN{{\cal N}}

\def\CP{{\cal P}}

\def\CW{{\cal W}}

\def\CZ{{\cal Z}}

\def\a{\alpha}
\def\b{\beta}
\def\g{\gamma}

\def\G{\Gamma}

\def\sc{\mathsf{c}}

\def\sst#1{{\scriptscriptstyle #1}}
\def\0{{\sst{(0)}}}
\def\1{{\sst{(1)}}}
\def\2{{\sst{(2)}}}
\def\3{{\sst{(3)}}}
\def\4{{\sst{(4)}}}
\def\5{{\sst{(5)}}}
\def\6{{\sst{(6)}}}
\def\7{{\sst{(7)}}}
\def\8{{\sst{(8)}}}

\newcommand{\be}{\begin{equation}}
\newcommand{\ee}{\end{equation}}
\newcommand{\bea}{\begin{eqnarray}}
\newcommand{\eea}{\end{eqnarray}}

\def\half{\frac{1}{2}}

\def\det{{\rm det}}

\def\CA{{\cal A}}

\def\CF{{\cal F}}

\def\CH{{\cal H}}

\def\CL{{\cal L}}

\def\CN{{\cal N}}

\def\CP{{\cal P}}

\def\CW{{\cal W}}

\def\CZ{{\cal Z}}

\def\a{\alpha}\def\b{\beta}\def\g{\gamma}

\def\G{\Gamma}

\def\a{\alpha}
\def\b{\beta}
\def\g{\gamma}

\def\G{\Gamma}

\def\sc{\mathsf{c}}

\def\sst#1{{\scriptscriptstyle #1}}
\def\0{{\sst{(0)}}}
\def\1{{\sst{(1)}}}
\def\2{{\sst{(2)}}}
\def\3{{\sst{(3)}}}
\def\4{{\sst{(4)}}}
\def\5{{\sst{(5)}}}
\def\6{{\sst{(6)}}}
\def\7{{\sst{(7)}}}
\def\8{{\sst{(8)}}}

\newcommand{\beq}{\begin{eqnarray}}
\newcommand{\eeq}{\end{eqnarray}}

\newcommand{\Li}{\ensuremath{\mathrm{Li}}}

\makeatletter
\def\Ddots{\mathinner{\mkern1mu\raise\p@
\vbox{\kern7\p@\hbox{.}}\mkern2mu
\raise4\p@\hbox{.}\mkern2mu\raise7\p@\hbox{.}\mkern1mu}}
\makeatother

\title{3d $\CN=2$   minimal SCFTs from Wrapped M5-branes}
\preprint{IPMU16-0085, KIAS-P16081}

\author[\epsdice{1}]{Jin-Beom Bae,}\author[\epsdice{2}]{Dongmin Gang,}
\author[\epsdice{3}]{and Jaehoon Lee}

\affiliation[\epsdice{1}]{School of Physics, Korea Institute for Advanced Study, \\
85 Hoegiro, Dongdaemun-Gu, Seoul 02455, Korea}
\affiliation[\epsdice{2}]{Kavli Institute for the Physics and Mathematics of the Universe (WPI),\\
University of Tokyo, Chiba 277-8583, Japan}
\affiliation[\epsdice{3}]{Department of Physics and Astronomy, University of British Columbia, \\
6224 Agricultural Road, Vancouver, BC, V6T 1W9, Canada}

\abstract{We study  CFT data of 3-dimensional superconformal field theories (SCFTs) arising from wrapped two M5-branes on closed hyperbolic 3-manifolds. Via so-called 3d/3d correspondence,   central charges of these SCFTs are related to a $SL(2)$ Chern-Simons (CS) invariant on the 3-manifolds. We  give a rigorous definition of the invariant  in terms of resurgence theory and a state-integral model for the  complex CS theory.  We numerically evaluate the central charges for several closed 3-manifolds with small hyperbolic volume. The computation suggests that the wrapped M5-brane systems  give  infinitely many discrete SCFTs with small central charges. We also analyze these `minimal' SCFTs  in the eye of 3d $\CN=2$ superconformal bootstrap.  
}

\begin{document}
\maketitle


\section{Introduction and Conclusion}

Quantum field theory (QFT) has become the dominant language in theoretical physics since the success of quantum electrodynamics.  The usage of QFT is not restrict to  particle physics but ubiquitous: statistical, condensed matter system and even quantum gravity using holography. 
In general, QFTs are in the form of
\begin{align}
\textrm{QFT} \;:\;  \textrm{(a CFT with flavor symmetry $G$)+(deformation)+(gauging $H\subset G$)}\;. \nn
\end{align}
At infrared (IR) limit, a QFT   flows to another conformal field theory (CFT). So, the general QFTs  can be thought as RG flows between CFTs and thus understanding general CFTs is the first step toward understanding  QFTs. 
 \begin{align}
\textrm{ `` Classify consistent CFTs and  solve them ''}   \nn
\end{align}
One rigorous way of defining a CFT is specifying CFT data: spectrum of local operators $\{ O_I\}$ and their operator product expansion (OPE) coefficients $\{\lambda_{IJ K}\}$. 
By solving a CFT, we mean determining these CFT data.

In this work, we study 3d $\CN=2$ unitary superconformal field theories (SCFTs) {\it without any flavor symmetry} and with {\it small central charges}.  3d supersymmetry has not been observed experimentally yet.  But there is a concrete proposal for condensed matter system~\cite{Ponte:2012ru,Grover:2013rc} which exhibits an emergent supersymmetry and  described by a 3d SCFT called  critical Wess-Zumino (cWZ) model. The model is known to be the simplest 3d $\CN=2$ SCFT with smallest central charge $c_T/c_T^{free} = \frac{16}{243} \left(16-\frac{9 \sqrt{3}}{\pi }\right)  \simeq 0.7268$ \cite{Witczak-Krempa:2015jca} where $c_T^{free}$ is the central charge for a free chiral theory. 
Classifying such simple unitary  CFTs  is an interesting open question.  In two dimensional spacetime, there is a complete classification when $c_T<1$ called  {\it 2d minimal models}.
Here, we propose 3d $\CN=2$ {\it `minimal' SCFTs} based on wrapped M5-brane systems.

\begin{figure}[h]
\begin{center}
   \includegraphics[width=1\textwidth]{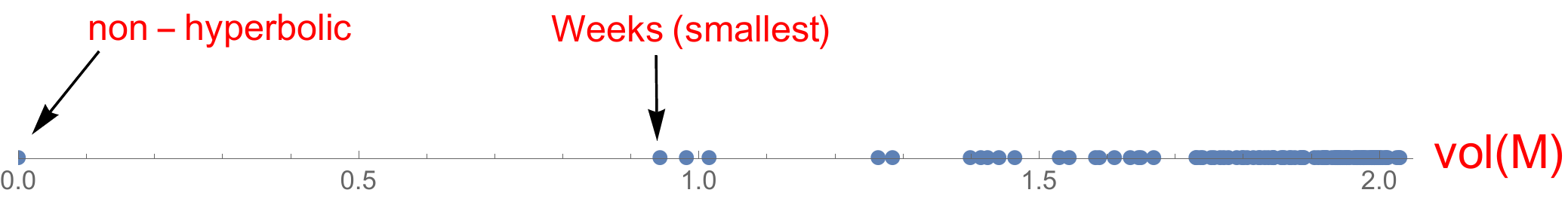}
   \end{center}
   \caption{Space of  closed   3-manifolds $M$ with $\textrm{vol}(M) <\textrm{vol}\big{(}(S^3\backslash {\bf 5^2_1})_{(5,-1)}\big{)}\simeq 2.02988$. ${\rm vol(M)}$  denotes a topological invariant of 3-manifold $M$ called {\it hyperbolic volume}, the volume measured in the unique  hyperbolic metric $(R_{\mu\nu}=-2g_{\mu\nu})$. For each non-zero hyperbolic volume plotted in the graph, there are only finitely many (mostly unique) ${\rm CH3}$s.
  The spectrum  is discrete and infinite   and has a (non-zero) lower bound 0.9427 which is saturated by  the {\it Weeks} manifold  \big{(}$(S^3\backslash {\bf 5^2_1})_{(5,-1)(5,-2)}$\big{)} \cite{2007arXiv0705.4325G}.  }   
   \label{fig:space of CH3}
\end{figure}

\begin{figure}[htbp]
\begin{center}
   \includegraphics[width=.2\textwidth]{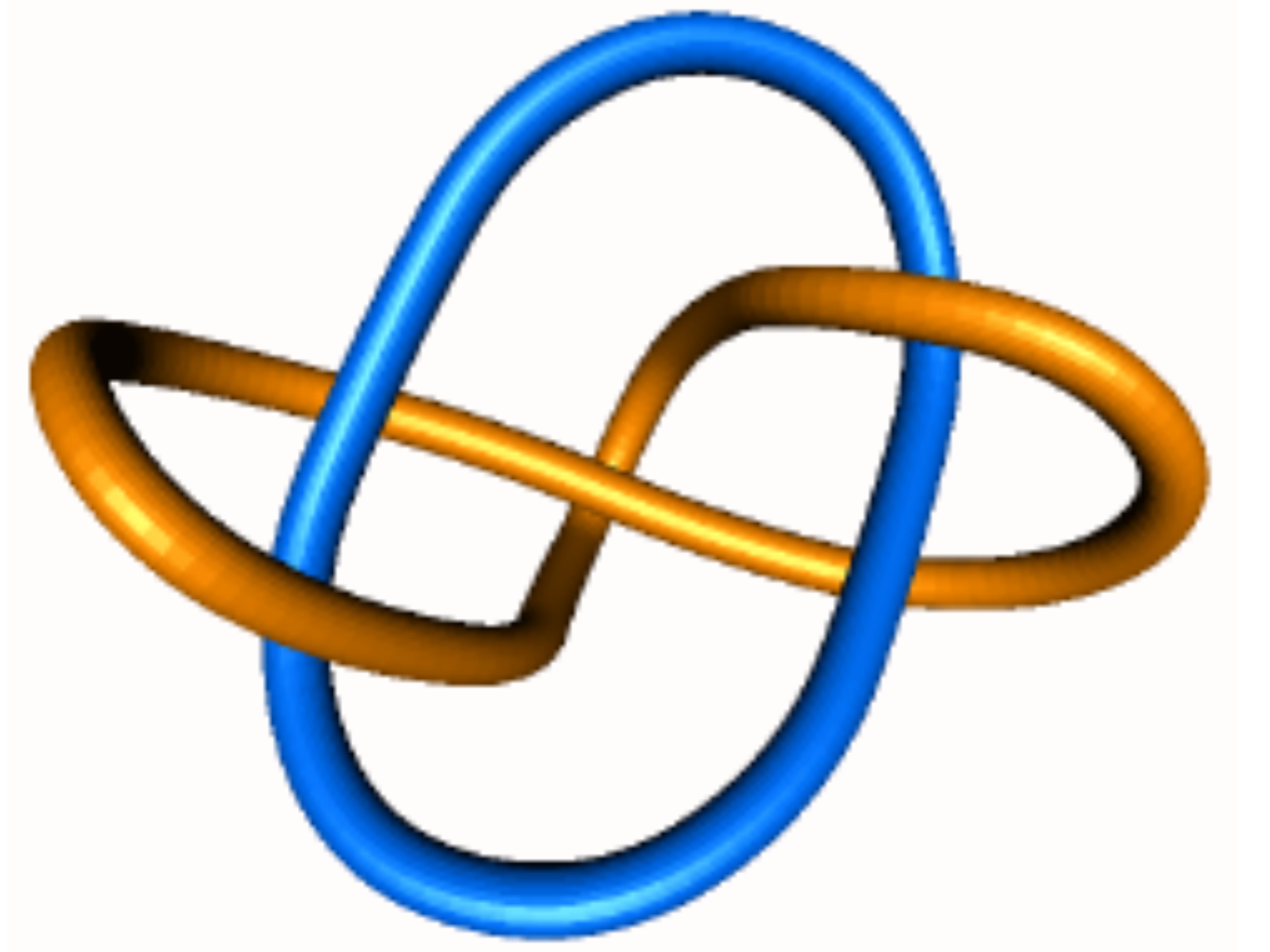}
   \end{center}
   \caption{White-head link (${\bf {\color{red}5}^{2}_{\color{blue}1}}$), the {\bf \color{blue}1}st one among links  with {\bf 2} components and {\bf \color{red}5} crossings.}
    \label{fig:White-head}
\end{figure}

 An efficient way of constructing  3d $\CN=2$  SCFTs is using wrapped M5-branes system in M-theory:  
\begin{align}
11d \; \textrm{space-time : }\;\; &\mathbb{R}^{1,2}\times T^*M \times \mathbb{R}^2\;  \nn
\\
&\;\;\; \;\; \;\;\bigcup \label{Wrapped M5-branes system}
\\
N \; M5\textrm{-branes}\; :\; \;\;& \mathbb{R}^{1,2}\times M\;. \nn
\end{align}
Here  $T^* M$ denotes the cotangent bundle of $M$.   The IR fixed point of the wrapped M5-branes' world-volume theory defines a 3d $\CN=2$ SCFT.   It is  labelled by an orientable closed hyperbolic 3-manifold (${\rm CH3}$) $M$ and an integer $N\geq 2$. We denote the SCFT  as $T_N[M]$.\footnote{For $N=2$ case,  we  skip the subscript ``$N$".}   The space of CH3 with small hyperbolic volume is depicted in Figure~\ref{fig:space of CH3}.  
 For  nomenclature of  3-manifolds, we use a Dehn surgery description \eqref{surgery description} on $S^3$ along  a White-head link  in Figure~\ref{fig:White-head}.  %
 One natural question is 
\begin{align}
\textrm{ `` Solve $T[M]$ for closed hyperbolic 3-manifolds $M$ ''.}  \label{solving wrapped M5-branes}
 \end{align}
 As a first step we develop a systematic algorithm for computing the central charge of wrapped M5-brane CFTs.   The algorithm can be summarized as :
\begin{align}
&\big{(} \textrm{a surgery description \eqref{surgery description} of $M$ and an ideal triangulation  \eqref{classical triangulation}} \big{)} \;,  \nn
\\
 &\xRightarrow{\;\;\;\;\;\;\;}  \big{(}\textrm{State-integral model in eq.~\eqref{quantum ideal triangulation} and \eqref{Dehns-filling}} \big{)}\;,   \nn
\\
& \xRightarrow{\;\; \;\;\;\;\;} \textrm{\big{(}Perturbative expansion $\{ S_n^{\overline{\textrm{hyp}}}(M)\}_{n=0}^{\infty}$ in  eq.~\eqref{def of S^{conj}}\big{)}}\;, \nn
\\
&\xRightarrow{\;\;\;\;\;\;\;} \textrm{\big{(}Borel resummation to $\CZ^{\overline{\textrm{hyp}}}(M;\hbar)$ in  eq.~\eqref{Z^{conj} from borel sum}\big{)}} \;, \label{All step in central charge computation}
\\
 &\xRightarrow{\;\;\;\;\;\;\;} \textrm{\big{(}$\CZ\big{[} \textrm{$T[M]$ on $S^3_b$}\big{]}$ using 3d/3d relation in eq.~\eqref{3d/3d in resurgence}\big{)}} \;, \nn
 \\
 &\xRightarrow{\;\;\;\;\;\;\;} \textrm{\big{(}$c_T \big{(}T[M] \big{)}$ using the relation in eq.~\eqref{squashed free-energy and central charge}\big{)}}\;. \nn
\end{align}
In the procedure, a) we reformulate the 3d/3d correspondence for squashed 3-sphere partition function (ptn) in the language of {\it resurgence} and b) develop a {\it state-integral} for $SL(2)$ Chern-Simons theory on {\it closed hyperbolic 3-manifolds}. We numerically evaluate the central charge for three examples  listed in Table~\ref{3 examples}.   %
\begin{table}[h]
\centering
\begin{tabular}{|c|c|c|c|}
\hline
 &{\it Weeks}: $(S^3\backslash {\bf 5^2_1})_{(5,-1)(5,-2)}$  & {\it Thurston}: $(S^3\backslash {\bf  5^2_1})_{(5,-1)(1,-2)}$ &  $(S^3\backslash {\bf 5^2_1})_{(5,-1)(5,-1)}$
 \\ \hline
 $c_T (T[M])$ &     0.93 & 1.01   &    1.28
 \\ \hline
 vol($M$) & 0.9427  &  0.9814&   1.2637
 \\ \hline
\end{tabular}
\caption{Central charge of $T[M]$ and hyperbolic volume for three CH3s.} \label{3 examples}
\end{table}
From the three examples, we see that the central charge is well-approximated by the hyperbolic volume within a few percent error. 
With this observation on top of {\it Weeks} manifold having smallest non-zero hyperbolic volume, we expect that the $T[{\it Weeks}]$ to be the simplest non-trivial wrapped M5-brane SCFT and there are infinitely many discrete  SCFTs with small central charge $(\lesssim 2)$. 

In recent years, the conformal bootstrap has provided valuable insights in strongly-interacting CFTs in $ d\geq2$ spacetime dimensions started with pioneering work of~\cite{Rattazzi:2008pe}.
Studying even a small subset of crossing symmetry constraints combined with unitarity was surprisingly restrictive on the allowed CFT data. 
The approach is quite universal and have been used to study CFTs in various spacetime dimensions with various number of supersymmetries and global symmetries~\cite{ElShowk:2012ht, El-Showk:2014dwa, Kos:2013tga, Nakayama:2014yia, Poland:2010wg, Poland:2011ey, Beem:2013qxa, Beem:2014zpa, Beem:2015aoa, Chester:2014fya, Iliesiu:2015qra, Lin:2015wcg, Lin:2016gcl}\footnote{See~\cite{Poland:2016chs} for a recent survey of conformal bootstrap for complete references.}.
It provides truly non-perturbative approach to probe full spaces of CFTs without reference to specific microscopic description, therefore providing another important tool for studying wrapped M5-brane CFTs.

For this purpose, we refer to the study for 3d $\CN=2$ was initiated in~\cite{Bobev:2015vsa,Bobev:2015jxa}.
In the dimension bound for unprotected operator there exist a kink which is connected to the kink observed in 4d $\CN=1$ superconformal bootstrap~\cite{Poland:2011ey, Poland:2015mta}.
It has been proposed that this kink might be signalling some unknown `minimal' SCFT\footnote{In 3d $\CN=2$ superconformal bootstrap, there are actually three kinks. cWZ model which appear in the first kink has smaller $c_T$ and is candidate for the global `minimal' theory. In the vicinity of the third kink, which we focus on this work there seems to be notion of local `minimality' at least respect to $c_T$.}. 
In this work, we further analyze and improve upon the results of~\cite{Bobev:2015vsa,Bobev:2015jxa} to make connections to wrapped M5-brane CFTs as possible candidates for the kink theory. 
Using chiral ring relation $\Phi^2 \sim 0 $ for chiral primary operator $\Phi$, our best estimate for the CFT data obtained using numerical superconformal bootstrap is
$\Delta_\Phi \geq 0.87$, $c_T/c_T^{free} \geq 1.0361$ and $\Delta_{\bar \Phi \Phi} \leq 2.3624$.

This work  put the first step toward the challenging problem \eqref{solving wrapped M5-branes} and there are  several interesting  directions worth exploring. We hope to report  progresses on these in near future.
\\
$\bullet$ Justify the physical  and technical  assumptions (\eqref{contour for real sigma} and \eqref{equivalence of two defs})  used in the central charge computation. We  give some circumstantial evidences for them.
\\
$\bullet$ Prove  topological  invariance  of the state-integral model developed in Section~\ref{sec: state-integral}. The state-integral model is based on a Dehn surgery representation \eqref{surgery description} of a 3-manifold. The representation is not unique and we need to show the independence on the choice.  We  check it perturbatively up to two-loops for several cases. 
\\
$\bullet$ Determine BPS operator spectrum  of the $T[M]$. As noticed in Section \ref{sec : BPS spectrum}, it is  related to  the problem of refinement/categorification  of Chern-Simons invariants on CH3s. 
\\
$\bullet$ Using  the central charge and BPS operator spectrum obtained above, bootstrap the $T[M]$  following \cite{Beem:2015aoa}.

The paper is organized as follows. In Section \ref{sec : wrapped M5 CFT}, we introduce wrapped M5-brane SCFTs and their basic properties. We  also present a rigorous form of 3d/3d relation for $S^3_b$-ptn in terms of resurgence theory. In Section \ref{sec : central charge},  a systematic algorithm for computing the central charge is given. It is based on  a state-integral  for a complex CS theory developed in the section. We give explicit examples for the computation and comments on the difficulties in determining chiral operator spectrum.  In Section \ref{sec: conformal bootstrap}, we investigate the possibility of SCFTs at  kinks in numerical bootstrap being identified with  wrapped M5-brane SCFTs.

\section{Wrapped M5-brane SCFT and 3d/3d correspondence} \label{sec : wrapped M5 CFT}
We introduce a 3d SCFT $T[M]$ labelled by a closed  3-manifold $M$ and  review basic aspects (holography and 3d/3d correspondence) of the SCFT. Especially, we reformulate the 3d/3d relation for squashed 3-sphere ptn in terms  of resurgence theory. For recent studies on the topic, refer to \cite{Dimofte:2010tz,Terashima:2011qi,Terashima:2011xe,Dimofte:2011jd,Dimofte:2011ju,Dimofte:2011py,Dimofte:2013iv,Gang:2013sqa,Yagi:2013fda,Lee:2013ida,Cordova:2013cea,Chung:2014qpa,Dimofte:2014zga,Gang:2014ema,Gang:2015wya,Pei:2015jsa} (see also recent review \cite{Dimofte:2016pua}). 

\subsection{3d $\CN=2$ SCFT $T_N[M]$}
A 3d SCFT $T_{N}[M]$ is defined as an infrared (IR) fixed point of twisted compactification of 6d $A_{N-1}$ $(2,0)$  theory on a closed 3-manifold $M$:
\begin{align}
&\textrm{6d $A_{N-1}$ (2,0) theory on $\mathbb{R}^{1,2}\times M$ with partial topological twisting along $M$}\; \nn
\\
& \xrightarrow{\;\;\;\;\textrm{IR}\;\;\;} \;\;\textrm{3d SCFT $T_{N}[M]$}\;. \label{RG : 6d -> 3d}
\end{align}
For the partial twisting, we use the usual $SO(3)$ subgroup of $SO(5)$ R-symmetry of the 6d theory. The twisting generically preserves  a quarter of supercharges and the resulting 3d theory has  $\CN=2$ superconformal symmetry.  The metric structure on the 3-manifold is irrelevant in  the IR  and the 3d SCFT  depends only on the topology of the 3-manifold. 
From M-theoretical perspective, these theories are realized as low-energy world-volume theory of wrapped   $N$ M5-branes in \eqref{Wrapped M5-branes system}.
As pointed out in \cite{Chung:2014qpa}, the `full' IR CFT   has an additional abelian flavor symmetry called $U(1)_t$ and  will be denoted as $T^{\rm full}_{N}[M]$. 
\begin{table}[h]
\centering
\begin{tabular}{|c|c|c|}
\hline
 & $T^{\textrm{full}}_{N}[M] $ &   $T_{N}[M]$    \\ \hline
 Global symmetry &    \;\; $U(1)_R\times U(1)_t$ & $ U(1)^{\rm IR}_R$      \\ \hline
 3d/3d correspondence & \;\;``Refined'' $SL(N)$ CS theory\;\; & $SL(N)$ CS theory    \\ \hline
Large $N$ gravity dual &  \textrm{Unknown}  &    $AdS_4\times M\times S^4$ (for $CH3$)\\ \hline
\end{tabular}
\caption{Comparison between $T^{\rm full}_{N}[M]$ and $T_N[M]$.} \label{Comparison between two wrapped M5 SCFTs}
\end{table}
The theory $T_{N}[M]$ of our interest is obtained as IR fixed point of  the $T^{\rm full}_{N}[M]$ through a renormalization group (RG) flow  triggered by a Higgsing/deformation procedure%
\begin{align}
\big{(}T^{\rm full}_{N}[M]+\textrm{ Higgsing/deformation} \big{)}\;\;  \leadsto \;\;T_{N}[M]\;. \label{T^{full}[M] to T[M]}
\end{align}
 Not all 3-manifolds $M$ give non-trivial interacting CFTs. Our basic assumptions are 
\\
\\
a) For hyperbolic 3-manifold $M$, the IR fixed theory $T_{N}[M]$ is non-trivial. 
\\
\\
b) For non-hyperbolic 3-manifold $M$ with $SO(3)$ Riemmanian holonomy (for example, $M=S^3$), the corresponding $T_N[M]$ seems to be  more or less trivial theories (theories only with topological degree of freedom).\footnote{The theory  $T^{\rm full}_{N}[M]$ might not be topological even this case. For example, $T^{\rm full}_{N}[S^3/\mathbb{Z}_p]$ is not topological \cite{Gaiotto:2007qi,Gukov:2015sna,Pei:2015jsa}.} 
\\
\\
c) $M$ has reduced Riemannina holonomy group (thus non-hyperbolic), i.e, $M=\Sigma \times S^1$ with a Riemann surface $\Sigma$. In the case, the resulting 3d SCFT has  additional structure, enhanced $\CN=4$  SUSY or  additional flavor symmetry.
\\
\\
Simple evidence for a)  is%
\begin{align}
&\lim_{b\rightarrow 0 } 2\pi b^2 \CF_b(T_{N}[M]) = \frac{N(N^2-1)}{6}\textrm{vol}(M) \;. \label{evidence}
\end{align}
Here  $\CF_b$ denotes the free-energy on a squashed 3-sphere $S^3_b$  \cite{Hama:2011ea},
\begin{align}
\CF_b (\textrm{a SCFT})&:=(\textrm{free-energy of the SCFT on $S^3_b$}) \nn
\\
&:= -\textrm{Re} \big{ (} \log \CZ[ \textrm{the SCFT on $S^3_b$}] \big{)}\;. \label{S^3_b free-energy}
\end{align}
Metrically, the curved background can be  realized as
\begin{align}
S^3_b = \{b^2|z|^2+b^{-2}|w|^2=1\;:\; (z,w)\in \mathbb{C}^2 \}\;, \quad \textrm{with real $b$}. \label{squashed 3-spherer}
\end{align}
The geometry has an exact symmetry exchanging $b\leftrightarrow b^{-1}$ and so does the free-energy $\CF_b$.
The  relation in eq.~\eqref{evidence} can be explained using a 3d/3d relation and  perturbative expansion of $SL(N)$ CS theory as we will see in the next section.
Since we are interested in a non-trivial  3d $\CN=2$ SCFT with small central charge and no extra structures (flavor symmetry or enhacencd SUSY), we concentrate on $N=2$ and the case a).  

\paragraph{Holographic dual}  Holographic  dual to the RG flow \eqref{RG : 6d -> 3d} across dimension  was   constructed in \cite{Gauntlett:2000ng}
\begin{align}
(\; AdS_7\times S^4\;\textrm{solution}\;)  \xrightarrow{\;\;\;\;\textrm{Holographic RG}\;\;\;} (\;\textrm{Pernici-Sezgin $AdS_4$ solution in } \;) \nn
\end{align}
and   M-theory on the $AdS_4$ solution  is proposed as  gravity dual of $T_{N}[M]$.  The supergravity solution is 
\begin{align}
AdS_4\times M\times S^4\;, \label{gravity dual of wrapped M5s}
\end{align}
with a warped product metric and the $S^4$ non-trivially fibred over the $M$ factor.
The supergravity solution  was found only for  closed hyperbolic $M$. From the holographic computation  using supergravity approximation, it has been predicted that \cite{Gang:2014ema}
\begin{align}
\lim_{N\rightarrow \infty} \frac{1}{N^3} \CF_{b} (T_{N}[M]) = \frac{(b+b^{-1})^2}{12\pi } \textrm{vol}(M) \;. \label{two evidences}
\end{align}
\subsection{3d/3d relation and resurgence  } Naively, 3d/3d relation relates the squashed 3-sphere ptn of $T_{N}[M]$ to ptn of  $SL(N)$ CS theory  on $M$.
\begin{align}
&\CZ[T_{N}[M]\textrm{ on }S^3_b] =\CZ \big{[}SL(N)_{k,\sigma} \textrm{ CS theory on }M \big{]} \nn
\\
&:= \int \frac{[D\CA]}{(\textrm{gauge})}  \exp \bigg{(}  \frac{i (k +\sigma)}{8\pi} \textrm{CS}[\CA] + \frac{i (k-\sigma)}{8\pi} \textrm{CS}[\bar{\CA}] \bigg{)} \;,  \label{3d-3d correspondence}
\end{align}
where $k$ and $\sigma$ are two coupling constants of the complex CS theory. $k\in \mathbb{Z}$ is a quantized CS level and the $\sigma$ can be either  real or purely imaginary. In the 3d/3d relation for $S^3_b$-ptn,  they are  \cite{Cordova:2013cea,Dimofte:2014zga}
\begin{align}
k=1\;\;  \textrm{and}\;\; \sigma= \frac{1-b^2}{1+b^2}\;.  \label{3d-3d correspondence-2}
\end{align}
$\CA,\bar{\CA}$ denote a pair of  $SL(N)$ gauge fields on $M$ and the CS functional is defined as
\begin{align}
\textrm{CS} [\CA] :=\int_M \textrm{Tr} \bigg{(} \CA \wedge d \CA +\frac{2}3 \CA \wedge \CA \wedge \CA  \bigg{)}\;.
\end{align}
The path-integral  on the complex CS theory is  ambiguous since there is no canonical choice for the path-integral contour and gauge transformation for  real $\sigma$. See sec 2.1 in \cite{Gang:2015wya} for discussion on the issue.

Goal of this section is   to make {\it  the 3d/3d relation more rigorous}  avoiding these ambiguities. For the purpose, we use the language of  {\it resurgence theory}. 

\paragraph{Perturbative ptn $\CZ^{\overline{\textrm{hyp}}}_{N;\rm pert}$}When $b^2\rightarrow 0^+$, the $S^3_b$-ptn has following asymptotic expansion \cite{Terashima:2011xe,Dimofte:2011jd}
\begin{align}
\CZ\big{[}T_{N}[M]\textrm{ on }S^3_b \big{]}   \xrightarrow{\;\;\;\;\textrm{$b^2 \rightarrow 0^+$}\;\;\;}  \sum_\alpha n_\alpha \CZ^{\a}_{N;\textrm{pert}} \big{(} M;\hbar\big{)} \;.
\end{align}
Here $\a$ labels $SL(N)$ flat connections on $M$ and $n_\a$ are  integer coefficients and $Z_{\textrm{pert}}^\a$ denotes the  formal perturbative expansion around the flat-connection $\CA^{\a}$ .
\begin{align}
\CZ_{N;\textrm{pert}}^\a (M;\hbar) := \exp \bigg{(} \frac{1}{\hbar}S^\a_0 (M;N) +S^\a_1 (M;N)+ \ldots+\hbar^{n-1}S^\a_n(M;N)+\ldots\bigg{)} \;. \label{perturbative CS ptn}
\end{align}
Through out the paper, we define
\begin{align}
\hbar := 2\pi ib^2 \in i \mathbb{R}_+\;.
\end{align}
$S^\a_n$ is the $n$-loop $SL(N)$ CS invariant on $M$. The classical part is 
\begin{align}
S^\a_0 = -\frac{1}2  \textrm{CS}[\CA^\a]\;. 
\end{align}
For hyperbolic 3-manifolds, there are two special flat connections, $\CA^{\textrm{hyp}}$ and $\CA^{\overline{\textrm{hyp}}}$, which can be constructed using the unique (complete) hyperbolic structure on $M$:
\begin{align}
\CA^{\textrm{hyp}}_N := \rho_N ( \omega+ i e) \;, \quad  \CA^{\overline{\textrm{hyp}}}_N :=\rho_N ( \omega- i e)\;, \label{A^hyp, A^conj}
\end{align}
where $e$ and $\omega$ are drei-bein and spin connection for the unique  hyperbolic structure respectively and $\rho_N$ is an embedding of $SL(2)$ into $SL(N)$ using the $N$-dimensional representation of $SL(2)\simeq SU(2)_\mathbb{C}$. Einstein equation with negative cosmology constant become flat connection equation through the above relation. Value of CS functional  for these flat connections are related to the hyperbolic volume of 3-manifold:
\begin{align}
\textrm{Im}\big{(}\textrm{CS}[\CA^{\textrm{hyp}}_N]  \big{)} = -\frac{1}3 N(N^2-1) \;\textrm{vol} (M)\;, \quad \textrm{Im}\big{(}\textrm{CS}[\CA^{\overline{\textrm{hyp}}}_N]  \big{)} = \frac{1}3 N(N^2-1) \;\textrm{vol} (M)\;.
\end{align}
These flat connections have  most exponentially growing and decaying    classical part $e^{\frac{1}\hbar S_0}$ when $b\in \mathbb{R}$ :
\begin{align}
\textrm{Im}\big{(}\textrm{CS}[\CA^{\textrm{hyp}}_N]  \big{)} <\textrm{Im}\big{(}\textrm{CS}[\CA^{\a}_N]  \big{)}  <\textrm{Im}\big{(}\textrm{CS}[\CA^{\overline{\textrm{hyp}}}_N]  \big{)} \;, \; \textrm{for any other flat-connections $\CA_N^\a$}. \label{two special properties of A^{conj},A^{hyp}}
\end{align}
From the  compatibility with the holographic prediction \eqref{two evidences} and an argument using a state-integral model,\footnote{The state-integral model  can be interpreted as an integral from localization for a SCFT, which can be identified as $T[M]$ \cite{Dimofte:2011ju}, if  one choose a proper converging integration contour of the form \eqref{geometric contour}.  For some knot complements, it's checked that the contour is homologically  equivalent to the steepest descendant contour (Lefschetz thimble) associated to the saddle point  in \eqref{the saddle point in state-integral} which corresponds to  the flat connection $ \CA^{\overline{\textrm{hyp}}}$. } it has been claimed that~\cite{Andersen:2011bt,Gang:2014ema}
\begin{align}
n_\alpha \neq  0 \; \textrm{ only for  $\a$ = $\overline{\textrm{hyp}}$} \;. \label{contour for real sigma}
\end{align}
It implies that the $S^3_b$-ptn is exponentially decaying  at small $b$ which seems to be an universal property of unitary non-topological 3d SCFTs.  Actually, the choice \eqref{contour for real sigma} with $n_{\overline{\textrm{hyp}}} =\pm 1$ maximizes the free-energy $\CF_b$ at small $b$, see  eq.~\eqref{two special properties of A^{conj},A^{hyp}}. We assume that this is the correct choice for the IR SCFT appearing in the 3d/3d relation. 

\paragraph{Borel resummation to $\CZ^{\overline{\textrm{hyp}}}_N$} The perturbative ptn $\CZ_{N;\textrm{pert}}^{\overline{\textrm{hyp}}}(M;\hbar)$ can be promoted to non-perturbative ptn through Borel resummation. For that, first reorganize the perturbative expansion in the following ways :
\begin{align}
\CZ^{\overline{\textrm{hyp}}}_{N;\textrm{pert}}(M;\hbar) = \exp \bigg{(} \frac{1}{\hbar} S_0^{\overline{\textrm{hyp}}}(M;N)+ S_1^{\overline{\textrm{hyp}}}(M;N) \bigg{)} \times \bigg{(} 1+\sum_{n=1}^\infty a^{\overline{\textrm{hyp}}}_n (M;N) (b^2)^n \bigg{)}\;, \nn
\end{align}
then the non-perturbative ptn $\CZ_N^{\overline{\textrm{hyp}}}$ is defined by Borel summation of the series $\{a_n^{\overline{\textrm{hyp}}}\}$:
\begin{align}
&\CZ^{\overline{\textrm{hyp}}}_N (M;\hbar):=   \exp \bigg{(} \frac{1}{\hbar} S_0^{\overline{\textrm{hyp}}}(M;N)+ S_1^{\overline{\textrm{hyp}}}(M;N) \bigg{)} \times \bigg{(}1+ \int_0^\infty d\zeta e^{-\frac{\zeta}{b^2} }  B^{\overline{\textrm{hyp}}}_N(\zeta) \bigg{)}\;,  \nn
\\
& \textrm{ where } B_N^{\overline{\textrm{hyp}}}(\zeta) := \sum_{n=1}^\infty  \frac{a^{\overline{\textrm{hyp}}}_n (M;N)}{(n-1)!} \zeta^{n-1} \;. \label{Z^{conj} from borel sum}
\end{align}
Here we assume that the series $\{ a_n^{\overline{\textrm{hyp}}} \}_{n=1}^\infty$ is Borel summable which is reasonable since the saddle point $\CA^{\overline{\textrm{hyp}}}$ gives the    smallest classical contribution  and  thus other saddle points can not appear as instanton trans-series. The $Z^{\overline{\textrm{hyp}}}$ is  determined by the perturbative invariants $\{S^{\overline{\textrm{hyp}}}_n (M)\}_{n=1}^\infty$, which  can be defined (with  mathematical rigour) and explicitly computed using  state-integral models, see eq.~\eqref{def of S^{conj}} for $N=2$.  Using the definition, the 3d/3d relation in eq.~\eqref{3d-3d correspondence} and \eqref{3d-3d correspondence-2} for hyperbolic 3-manifolds $M$ can be more rigorously stated as:
\begin{align}
\CZ\big{[}T_N[M] \textrm{ on $S^3_b$}\big{]}  = \CZ_N^{\overline{\textrm{hyp}}}\big{(}M;\hbar=2\pi i b^2 \big{)}\;. \label{3d/3d in resurgence}
\end{align}
On the other hand, it was claimed in~\cite{Gukov:2016njj} that the Borel resummation $\CZ_N^{\overline{\textrm{hyp}}}$ gives the vortex ptn (ptn on $\mathbb{R}^2\times_q S^1$) instead of $S^3_b$-ptn.  There are two  evidences supporting our proposal over their claim: a) At large $N$ and the leading order $(N^3)$ in $1/N$ expansion, the perturbative series $\{S^{\overline{\textrm{hyp}}}_n (N)\}$ becomes a finite series terminating at two-loops and the  answer nicely matches with the holographic prediction \eqref{two evidences} of $S^3_b$-ptn \cite{Gang:2014ema}, b) For $N=2$ and $M=S^3\backslash \mathbf{4}_1$ (figure-eight knot complement), the  Borel resummation is performed  explicitly in  \cite{Gukov:2016njj} \footnote{There  seems to be a mistake in the sign of classical part in the eq. (6.11) in  \cite{Gukov:2016njj}.  After correcting the mistake, $\CZ^{\overline{\textrm{hyp}}}\big{(}S^3\backslash \mathbf{4}_1;\hbar=2\pi i b^2 \big{)}\big{|}_{b=1}= e^{-2 \times \frac{ \textrm{vol}(S^3\backslash \mathbf{4}_1)}{2\pi}} \times$\big{(}eq. (6.23) in \cite{Gukov:2016njj}\big{)}. }
\begin{align}
\CZ^{\overline{\textrm{hyp}}}\big{(}S^3\backslash \mathbf{4}_1;\hbar=2\pi i  b^2 \big{)}\big{|}_{b=1} \simeq 0.37953\;, \label{figure-eight resurgence}
\end{align}
which is a good approximation for the correct $S^3_b$-ptn of $T_{N=2}[S^3\backslash \mathbf{4}_1]$ computed using a state-integral model. The exact value at $b=1$ is  \cite{Garoufalidis:2014ifa}
\begin{align}
\CZ\big{[}T_{}[S^3\backslash \mathbf{4}_1] \textrm{ on $S^3_{b=1}$}\big{]}  &=\frac{1}{\sqrt{3}} \bigg{(}\exp(\frac{\textrm{vol}(S^3\backslash \mathbf{4}_1)}{2\pi})-\exp(-\frac{\textrm{vol}(S^3\backslash \mathbf{4}_1)}{2\pi}) \bigg{)} \simeq 0.379568\;.  \label{figure-eight state-integration}
\end{align}
Here the hyperbolic volume of $S^3\backslash \mathbf{4}_1$ is
\begin{align}
\textrm{vol}(S^3\backslash \mathbf{4}_1) = 2 \textrm{Im}[\Li_2 (e^{i \pi/3})] \simeq 2.02988\;.
\end{align}
The proposed equality \eqref{3d/3d in resurgence} is somewhat surprising since the quantity in the left-hand side has a manifest $b\leftrightarrow b^{-1}$ symmetry but the other does not. In the asymptotic expansion, $S_n^{\overline{\textrm{hyp}}}$ or $a^{\overline{\textrm{hyp}}}_n$, the non-perturbative symmetry is invisible but the equality suggests that {\it the symmetry  emerges after Borel resummation}. %
It would be interesting to explicitly check the emergence for several examples. 
\section{CFT data of $T[M]$}
\label{section:CFTdata}  \label{sec : central charge}

\subsection{Central charge computation}
One basic quantity characterizing a SCFT is central charge $c_T$ which is  defined using two point function of stress-energy tensor: 
\begin{align}
T(x)T(0) \sim \frac{c_T}{|x|^{2d}} \times (\textrm{tensor structure}) \;.
\end{align}
For 3d $\CN=2$ SCFTs, the central charge is related to the squashed 3-sphere free energy $\CF_b$ \eqref{S^3_b free-energy} as follows  \cite{Closset:2012ru}:
\begin{align} 
&c_T =  \frac{8}{\pi^2} \frac{\partial^2 \CF_b}{\partial b^2}\bigg{|}_{b=1}   \;. \label{squashed free-energy and central charge}
\end{align}
We use following normalization 
\begin{align}
c_T (\textrm{a free chiral theory})  = 1\;. \label{normalization of central charge}
\end{align}
Combining the 3d/3d correspondence \eqref{3d/3d in resurgence} and the relation \eqref{squashed free-energy and central charge}, we will compute the central charge  of $T[M]$. The full procedure is summarized in eq.~\eqref{All step in central charge computation}. 

\subsubsection{A state-integral model for $SL(2)$ CS theory} \label{sec: state-integral}
As a first step in \eqref{All step in central charge computation}, we review and extend a state-integral for $SL(2)$ CS theory on hyperbolic 3-manifolds which gives a  rigorous definition and a computation tool for the CS perturbative invariants $\{S^{\overline{\textrm{hyp}}}_n(M)\}_{n=0}^{\infty}$. The extended state-integral model {\it is  applicable to any closed hyperbolic 3-manifolds} which was not possible for  state-integrals  \cite{Dimofte:2011gm,Hikami01,Andersen:2011bt} in the literature. 

\paragraph{Dehn surgery and ideal triangulation} We  use a Dehn surgery description of 3-manifold $M$ :%
\begin{align}
&M= \big{(}S^3\backslash K\big{)}_{\{(p_\a,q_\a)\}_{\a=1}^{S\leq |K|} }:= \bigg{[}\big{(}S^3 \backslash K \big{)} \bigcup_{\a=1}^S (D^2\times S^1)_\a \bigg{]} /\sim \;, \label{surgery description}
\end{align}
and  a sufficiently good\footnote{At least, we assume   a positive angle structure of  triangulation \cite{Dimofte:2014zga}.} ideal triangulation of  the link complement $S^3\backslash K$ :
\begin{align}
S^3\backslash K = \bigg{(}\bigcup_{i=1}^T \Delta_i \bigg{)}/\sim\; . \label{classical triangulation}
\end{align}
  \begin{figure}[h]
\begin{center}
   \includegraphics[width=.25\textwidth]{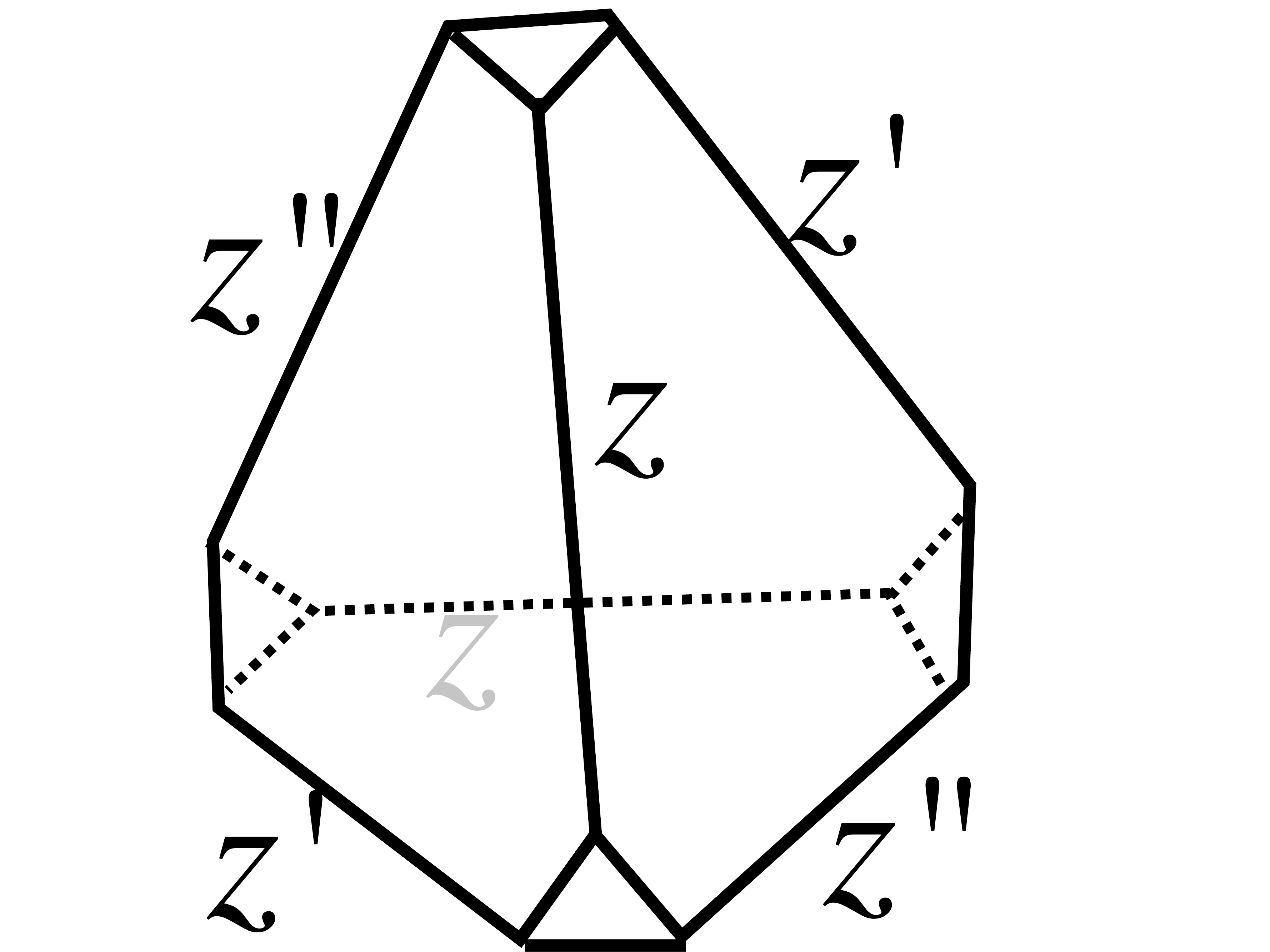}
   \end{center}
   \caption{ An ideal tetrahedron $\Delta$, tetrahedron with truncated vertices. Hyperbolic structures of $\Delta$ are parameterized by edge parameters $(z:=e^{Z},z':=e^{Z'},z''=e^{Z''})$ satisfying relations $z'=\frac{1}{1-z}$ and $z''=1-z^{-1}$.  These parameters assigned to each pair of boundary edges, as shown in the figure.  Geometrically,  the logarithm parameters ($Z,Z',Z''$) measure  complex dihedral angles between two faces meeting at the edges. Imaginary parts of these logarithm parameters take values  between 0 and $\pi$.   }   
   \label{fig:ideal-tetrahedron}
\end{figure}
Here $K$ is a  link on $S^3$ of $|K|$ components.  A link complement $S^3\backslash K$ is a 3-manifold obtained by removing the tubular neighborhood (topologically $|K|$ copies of solid-tori) of a link $K$ from a 3-sphere $S^3$.
The manifold has $|K|$ torus boundaries and 1-cycles around the link are called `meridians' and 1-cycles along the link are `longitudes'. The 3-manifold  $M$ in \eqref{surgery description} is obtained by gluing $S$ solid-tori back to the link complement with following identification :
 \begin{align}
&p_\a (\textrm{$\a$-th meridian})+q_\a( \textrm{$\a$-th longitude}) \nn
\\
 &\sim (\textrm{contractable cycle  in $\a$-th solid-torus})\;. \label{classical Dehn's filling}
\end{align}
The procedure of gluing solid-torus is called $(p_\a, q_\a)$-Dehn filling.
$(p_\a,q_\a)$  is a pair of coprime numbers  and the ratio $p_\a /q_\a$ is called `slopes'. 
In short, the 3-manifold is obtained by gluing $T$ ideal tetrahedrons and $S$ solid-tori:
\begin{align}
T\; :\; \sharp \textrm{ of ideal tetrahedrons}\;, \quad S \;:\;\sharp \textrm{ of solid-tori}\;.
\end{align}
The resulting 3-manifold $M$ has $(|K|-S)$ torus boundaries and when $S=|K|$ it is  a closed 3-manifold.  {\it Any closed 3-manifold $M$ can be obtained by  a Dehn surgery on $S^3$} \cite{10.2307/1970373,Wallace1960}.

\paragraph{State-integral model } State-integrals  give a finite-integral  representation of the CS ptn by properly `quantizing' the ideal triangulation \eqref{classical triangulation} and the Dehn filling \eqref{classical Dehn's filling}. There are several state-integral models \cite{2007JGP....57.1895H,Dimofte:2011gm,Andersen:2011bt}, which are believed to be equivalent, based on an ideal triangulation of $M$. We  use the one developed by Dimofte and incorporate Dehn filling into the state-integral model to {\it cover more general class of 3-manifolds such as closed hyperbolic 3-manifolds}. One systematic way of specifying the gluing rule of an ideal triangulation is using (generalized) Neunmann-Zagier (NZ) datum $(A,B,C,D;f,f'',\nu,\nu_p)$, refer to \cite{Dimofte:2012qj} for the definition, where $A,B,C,D$ are $T\times T$ matrices forming $Sp(2T,\mathbb{Q})$
 \begin{align}
 \left(\begin{array}{cc}A & B \\C & D\end{array}\right) \in Sp(2T,\mathbb{Q})\;, \quad \textrm{with } \det B\neq 0\;, 
 \end{align}
 and $(f,f'',\nu,\nu_p)$ are   vectors of length $T$. From these datum, the state-integral ({\rm SI}) for the link complement is given by \cite{Dimofte:2012qj}
\begin{align}
&\CZ_{\rm SI} (S^3\backslash K;X_1, \ldots, X_{|K|};\hbar) 
\nn
\\
&= \frac{1}{\sqrt{\det B}}\int \prod_{i=1}^T \frac{ \Psi_b (Z_i) dZ_i}{\sqrt{2 \pi \hbar}}  \exp \bigg{[}  \frac{1}{2  \hbar} \vec{Z}\cdot B^{-1} A \vec{Z} +\frac{1}{\hbar} \bigg{(} 2 u \cdot D B^{-1} u+  (2\pi i +\hbar) f\cdot B^{-1} u  \nn
\\
&\qquad \qquad \qquad \qquad +\half (i \pi + \hbar/2)^2 f\cdot B^{-1} \nu - \vec{Z}\cdot B^{-1} \big{(}(i \pi +\hbar/2) \nu+2 u\big{)} \bigg{)} \bigg{]}   \;.\label{quantum ideal triangulation}
\end{align}
\\
Here we define
\begin{align}
u = \big{(}X_1,\ldots, X_{|K|},0,\ldots, 0 \big{)} \;. \label{def :u}
\end{align}
The quantum dilogarithm function (QDL) $\Psi_b$ is a wave-function on each tetrahedron. See Appendix~\ref{App : QDL} for the definition and basic properties of the special function. Quantizing the Dehn fillings in \eqref{classical Dehn's filling}, we finally have
\begin{align}
&\CZ_{\rm SI} \big{(} M;X_1, \ldots, X_{|K|-S};\hbar\big{)} 
\nn
\\
&=\int \prod_{\a=1}^{S}\frac{\Delta_b(X_{\a+|K|-s};s_{\a},q_{\a}) dX_{\a+|K|-S}}{(2 \pi q_{\a} \hbar )^{1/2}}  \exp \bigg{(}\frac{1}\hbar \sum_{\a=1}^{S}\frac{p_\a X_{\a+|K|-s}^2} {  q_\a}  \bigg{)} \CZ_{\rm SI} \big{(} S^3\backslash K;X_1, \ldots, X_{|K|} ;\hbar \big{)}\;,  \nn
\\
&  \textrm{with }\Delta_b (X;s,q):= e^{- \frac{i \pi s}{2q}(b^2+b^{-2})} \bigg{(}e^{\frac{X}{b^2 q}} \sinh(\frac{X-i \pi s}q) -e^{-\frac{X}{b^2 q}}
 \label{Dehns-filling}
\sinh(\frac{X+i \pi s}q)\bigg{)} \;.\end{align}
Here $s_\a$ is defined to be an integer satisfying $ s_\a p_\a  \in q_\a \mathbb{Z} -1$. See Appendix~\ref{App : Dehn's filling} for the derivation. The CS wave-function has following naive path-integral interpretation,
\begin{align}
&\CZ_{\rm SI} \big{(} M;X_1, \ldots, X_{|K|-S};\hbar=2\pi ib^2\big{)}  \nn
\\
&  = \int \frac{[d\mathcal{A}]_X}{(\textrm{gauge})} \exp \bigg{(}  \frac{i (k+\sigma) }{8\pi } \textrm{CS}[\CA] + \frac{i  (k-\sigma)}{8\pi} \textrm{CS}[\bar{\CA}] \bigg{)}\bigg{|}_{\eqref{3d-3d correspondence-2}}\;, \;\textrm{  where}\nn
\\
& [d\CA]_X \; :\; \textrm{Path-integral over $SL(2)$ gauge field on $M$ subject to} \nn 
\\
& \qquad  \qquad \textrm{boundary conditions fixing }\textrm{P} e^{ \oint_{I-\textrm{th merdian}} \CA} = \left(\begin{array}{cc}e^{X_I} & 1 \\0 & e^{-X_I}\end{array}\right) \;. \label{boundary fixing conditions}
\end{align}
The $SL(2)$ CS wave-function is defined up to a  factor  \cite{Dimofte:2012qj}. 
\begin{align}
\exp \big{(}\frac{\pi^2}{6\hbar} \a + \frac{ i \pi }4  \b + \frac{\hbar}{24} \g \big{)}\;, \quad \a,\b,\g \in \mathbb{Z}\;. \label{intrinsic ambiguity}
\end{align}
The factor  is a purely phase factor for real $b$ and irrelevant in  free-energy $\CF_b$ computation. 
In the SCFT side of 3d/3d correspondence, (some parts of) the ambiguities  comes from   local counter-terms in a supergravity on the curved ($S^3_b$) background \cite{Closset:2012vg}. 

\subsubsection{Perturbative invariants } 
Using the state-integral model above, we  define the perturbative invariants which play an essential role in the 3d/3d relation \eqref{3d/3d in resurgence}. The state-integral model in \eqref{quantum ideal triangulation} and \eqref{Dehns-filling}   is of the form :
\begin{align}
&\CZ_{\rm SI} \big{(} M;X_1, \ldots, X_{|K|-S};\hbar\big{)}  \nn
\\
&= \int \frac{d X_{|K|-S+1} \ldots dX_{|K| } dZ_1 \ldots d Z_T}{(2\pi \hbar)^{(T+S)/2}} \; \exp \bigg{(}  \CW  \big{(}Z_1, \ldots, Z_T, X_1, \ldots, X_{|K|};\hbar \big{)} \bigg{)}\;. \label{form of state-integral model}
\end{align}
In the limit when $\hbar\rightarrow 0$, using eq.~\eqref{asymptotic of QDL}
\begin{align}
\CW (\vec{Z},\vec{X};\hbar) \sim \frac{1}{\hbar}\CW_0 (\vec{Z},\vec{X})  + \CW_1 (\vec{Z},\vec{X})+ \hbar \CW_{2}(\vec{Z},\vec{W})+\ldots \;.
\end{align}
Saddle point equations are
\begin{align}
&\bullet \; \frac{\partial \CW_0}{ \partial Z_i} =  0 \; , \quad \textrm{for } i=1,\ldots,T \nn
\\
&\;\; \;\;\Rightarrow 
A\cdot \vec{Z} +B \cdot \vec{Z}''- i \pi \nu =2 u\;\; \textrm{where } Z''_i := \log (1-e^{-Z_i})\;, \nn
\\ 
&\bullet \;\frac{\partial \CW_0}{ \partial X_{\a+|K|-S}} =  0\;,  \; \;   \textrm{for }\a=1, \ldots, S \nn
\\
&\;\; \;\;\Rightarrow  p_\a X_{\a +|K|-S} + q_\a P_{\a+|K|-S}=   -\textrm{sign} \big{(}\textrm{Re}[ \frac{X_{\a+|K|-S}}{q_\a}] \big{)} \pi i  \;.  \label{saddle point equations} 
\end{align}
Here $u$  is defined in \eqref{def :u} and  we define
\begin{align}
P_{\a+|K|-S}:=\big{(}C \cdot \vec{ Z}+ D \cdot \vec {Z}'' - i \pi \nu_p \big{)}_{\a+|K|-S}\;.
\end{align}
Interpreting  the variables $Z$ and $Z''$ as logarithmic edge parameters of ideal tetrahedrons, these  are nothing but gluing equations for the 3-manifold studied in \cite{NZ:1985}. Solutions to the gluing solution give  $SL(2)$ flat connections on $M$.
Refer to \cite{Dimofte:2013iv} for   explicit construction of holonomy  representation of a  flat connection from a solution to the gluing equations. In the map,  the solution  corresponding to the flat connection $\CA^{\overline{\rm hyp}}$  is characterized by following conditions:
\begin{align}
&0<\textrm{Im}[Z_i]<\pi\;, \quad \textrm{for all $i=1,\ldots, T$\;}\;\;\quad \quad(\textrm{hyperbolic}) \nn
\\
&X_1= \ldots = X_{|K|-S}=0\; \qquad \qquad \qquad \qquad  \;\;\;  (\textrm{complete}) \label{geometric gluing solution}
\end{align}
Under the first condition, logarithmic  edge parameter $Z_i$ determines a hyperbolic structure on $\Delta_i$, see  Fig \ref{fig:ideal-tetrahedron}. The gluing equations are  conditions for the hyperbolic structures  to be glued smoothly and give a hyperbolic structure on the 3-manifold. For  {\it complete} hyperbolic structure, we additionally need the second conditions requring  the meridian holonomies in eq.~\eqref{boundary fixing conditions} are parabolic.
Near each $\mathbb{T}^2$-boundary, the complete hyperbolic metric on $M$ are locally
\begin{align}
ds^2 = \frac{1}{z^2}\big{(} dz^2+ds^2_{\mathbb{T}^2} \big{)}\;.
\end{align}
Here $z$ is  the (inward)  direction transverse to the boundary $\mathbb{T}^2$. Using the metric, one can check that the  flat connection $\CA^{\overline{\textrm{hyp}}}$  in  \eqref{A^hyp, A^conj}  have parabolic meridian holonomies.
For the case when $M$ is hyperbolic and we use an idea triangulation with positive angle structure,  there is an unique  solution for eq.~\eqref{saddle point equations} and \eqref{geometric gluing solution} modulo the Weyl-symmetries $(\mathbb{Z}_2)^S$.
\begin{align}
 (\mathbb{Z}_2)^S \;:\; X_{\a+|K|-S}  \rightarrow \pm X_{\a+|K|-S} \;\; \; \textrm{for }\a=1,\ldots,S\;.  \label{Weyl-symmetry}
 \end{align}
The unique saddle point corresponds to the flat connection $\CA^{\overline{\textrm{hyp}}}$ and we denote
\begin{align}
\big{(} X_{\a+|K|-S}^{\overline{\textrm{hyp}}}, Z_i^{\overline{\textrm{hyp}}} \big{)} :=\textrm{A solution satisfying eq.~\eqref{saddle point equations} and \eqref{geometric gluing solution}} \;. \label{the saddle point in state-integral}
\end{align}
For non-hyperbolic $M$, there's no saddle point satisfying these conditions. The formal  perturbative expansion of the state-integral around the saddle point defines  the perturbative ptn $\CZ_{\rm pert}^{\overline{\textrm{hyp}}}(M;\hbar)$  \eqref{perturbative CS ptn}:
\begin{align}
&\CZ^{\overline{\textrm{hyp}}}_{\rm pert} \big{(}M;\hbar \big{)}  := 2^S \times \CZ_{\rm pert;SI}^{\overline{\textrm{hyp}}} \big{(}M;\vec{X}=\vec{0};\hbar \big{)}  \;, \nn
\\
&:= 2^S\times \big{[}\textrm{Perturbative expansion of $\CZ_{\rm SI}  \big{(}M;\vec{X}=\vec{0};\hbar \big{)}$ around \eqref{the saddle point in state-integral}} \big{]}\;.
 \label{def of S^{conj}}
\end{align}
The overall factor $2^{S}$ comes from the fact that there are that many saddle points related by Weyl-symmetries and they all give same perturbative expansion.
The state-integral is finite dimensional integration and thus the formal expansion coefficients $\{S^{\overline{\textrm{hyp}}}_{n}(M)\}_{n=0}^\infty$ are well-defined without any issue of regularization. Refer to \cite{Dimofte:2012qj} for perturbative expansion  of the state-integral model in \eqref{quantum ideal triangulation} using Feynman diagram.

 \paragraph{Examples  $(S^3\backslash {\bf 5^2_1})_{(p,q)}$}  White-head link (${\bf 5^2_1}$) is one of simplest hyperbolic link with two components. 
\\
 The link complement  can be decomposed into 4 ideal tetrahedrons (see Appendix  \ref{App : NZ-datum}) :
\begin{align}
S^3\backslash {\bf 5^2_1}   =\bigg{(}\bigcup_{i=1}^4 \Delta_i \bigg{)}/\sim  \;.
\end{align}
Using the ideal triangulation, the corresponding state-integral is given by
\begin{align}
&\CZ_{\rm SI} \big{(}S^3\backslash {\bf 5^2_1} ;X_1,X_2;\hbar \big{)}= \frac{1}{\sqrt{2}}  \int \prod_{i=1}^4 \frac{\Psi_b(Z_i) dZ_i}{\sqrt{2\pi \hbar}} \exp \bigg{[}\frac{2X_1 (2 Z_1+2 Z_4- \hbar -2 i \pi )-2 Z_3 Z_4}{2\hbar} \nn
\\
&\;\;\;\; \quad  +\frac{2 X_2 (-2 Z_2-2 Z_4+\hbar +2 i \pi )+(Z_1+Z_2-Z_3) (Z_1+Z_2-Z_3-\hbar -2 i \pi )}{2 \hbar } \bigg{]}\;.
\end{align}
Applying the quantum Dehn filling formula  \eqref{Dehns-filling} to the above integral, we obtain the state-integral for $M=(S^3\backslash {\bf 5^2_1})_{(p,q)}$. For example, when $(p,q)= (5,-1)$
\begin{align}
&\CZ_{\rm SI} \big{(} (S^3\backslash {\bf 5^2_1})_{(5,-1)} ;X_1;\hbar\big{)}  \nn
\\
&=  \int  \frac{ 2 \sinh (X_2)\sinh(X_2/b^2)dX_2}{\sqrt{2\pi \hbar}} \exp \big{(}-\frac{5 X_2^2}{\hbar}\big{)} \CZ_{\rm SI} (S^3\backslash {\bf 5^2_1} ;X_1,X_2;\hbar)\;. \label{integrand (5^2_1)_(5,-1)}
\end{align}
In the case, the resulting  3-manifold is turned out to be a 3-manifold called `sister of figure-eight knot-complement'. In  {\tt SnapPy}'s notation \cite{snappy}, the 3-manifold is denoted as $m003$ and allows an ideal triangulation using two tetrahedrons (see Appendix  \ref{App : NZ-datum}):
 \begin{align}
 m003 = (S^3\backslash {\bf 5^2_1})_{(5,-1)}= \bigg{(} \bigcup_{i=1}^{2}\Delta_i \bigg{)}/\sim \;.
 \end{align}
From the ideal triangulation, we have an alternative expression for the state-integral model
\begin{align}
\CZ_{\rm SI} (m003 ;X_1;\hbar) &=  \int \prod_{i=1}^2 \frac{\Psi_b(Z_i) dZ_i}{\sqrt{2\pi \hbar}} \exp \bigg{[}\frac{X_1 (8 Z_1+4 Z_2-2 \hbar -4 i \pi )+8 X_1^2}{2\hbar} \nn
\\
&\qquad \qquad  +\frac{2 Z_1 (Z_2-\hbar -2 i \pi )+Z_2 (Z_2-\hbar -2 i \pi )+4 Z_1^2}{2 \hbar } \bigg{]}\;.\label{m003-from-ideal}
\end{align}
One can check that both expressions, eq.~\eqref{integrand (5^2_1)_(5,-1)} and eq.~\eqref{m003-from-ideal}, give same perturbative invariants $S^{\overline{\textrm{hyp}}}_{n}(M)$ modulo \eqref{intrinsic ambiguity} :
\begin{align}
&S^{\overline{\textrm{hyp}}}_0 ( m003 )= S^{\overline{\textrm{hyp}}}_0 \big{(}(S^3\backslash {\bf 5^2_1})_{(5,-1)} \big{)} =(1.64493\, -2.02988 i)\;, \nn
\\
&  S^{\overline{\textrm{hyp}}}_1 ( m003 )= S^{\overline{\textrm{hyp}}}_1 \big{(}(S^3\backslash {\bf  5^2_1})_{(5,-1)} \big{)}=(-0.621227+1.309 i)\;, \nn
\\
&S^{\overline{\textrm{hyp}}}_2 ( m003 )=S^{\overline{\textrm{hyp}}}_2 \big{(}(S^3\backslash {\bf 5^2_1})_{(5,-1)} \big{)}=(0.0104167\, +0.0701641 i)\;. \nn
\end{align}
We did similar consistency checks for  other examples, $(S^3\backslash {\bf 5^2_1})_{(3,-2)} = m007,(S^3\backslash{\bf  5^2_1})_{(5,-2)} = m006$ and $(S^3\backslash {\bf 5^2_1})_{(2,-3)} = m053$. The matches are  delicate and  strongly suggests that  the state-integral model gives at least the correct  perturbative invariants. We leave  the general proof showing  topological invariance of the perturbative series as future work.

\subsubsection{Examples : $M=(S^3\backslash {\bf 5^2_1})_{(5,-1),(p,q)}$}
Here we give concrete examples of  central charge computation for closed hyperbolic 3-manifolds $M$.
The most technically non-trivial step in \eqref{All step in central charge computation} is extracting the perturbative invariants $\{ S_n^{\overline{\textrm{hyp}}}\}_{n=0}^\infty$ from the state-integral model and performing their Borel resummation. Since the central charge is related to the squashed 3-sphere ptn around $b=1$, which corresponds to $\hbar = 2\pi i = o(1)$, we need to take into account of sufficiently higher loop corrections to give a valid approximation.  To circumvent the difficulty, we  use a following alternative definition  of $\CZ^{\overline{\textrm{hyp}}}$:
\begin{align}
&\CZ^{\overline{\textrm{hyp}}} (M;\hbar)  \nn
\\
&:= \big{(}\textrm{Integration of the  $\CZ_{\rm SI}(M;\vec{X}=\vec{0};\hbar)$ in  \eqref{form of state-integral model} along a Contour $\Gamma^{\overline{\textrm{hyp}}} (b)$}\big{)}\;, \label{Z^{conj} from state-integral}
\end{align}
where the converging continuous Contour $\G^{\overline{\textrm{hyp}}}(b)$
\begin{align}
&\G^{\overline{\textrm{hyp}}}(b) :=  \big{\{} Z_i =m_i +i \mathbb{Z}_i (b;\vec{m}),\;  X_{|K|-S+\a} = m_{T+\a} +i  \mathbb{X}_\a (b;\vec{m})\;:\; \vec{m} \in \mathbb{R}^{T+S} \big{\}}_{i=1,\ldots,T}^{\a=1,\ldots, S}  \nn
\\
&\qquad \qquad \;\; \subset \mathbb{C}^{T+S}\;, \nn
\end{align}
 is chosen to satisfy following conditions
\begin{align}
&0<\mathbb{Z}_i (b;\vec{m})<\pi (1+b^2)\;,\quad \textrm{ $\forall \vec{m} \in \mathbb{R}^{T+S}$ and $\forall i=1,\ldots ,T$\;.}
\label{geometric contour}
\end{align}
From  Picard-Lefschetz  and resurgence theory (see, for example, \cite{Dunne:2015eaa} and reference therein) and the uniqueness of  saddle point \eqref{the saddle point in state-integral}, we expect that  for hyperbolic $M$
\begin{align}
&a) \textrm{ There is an unique converging non-trivial contour satisfying the conditions in \eqref{geometric contour} ,} \nn
\\
& b) \textrm{ Two definitions of $\CZ^{\overline{\textrm{hyp}}}$ in \eqref{Z^{conj} from borel sum} and \eqref{Z^{conj} from state-integral} are equivalent. } \label{equivalence of two defs}
\end{align}
For non-hyperbolic $M$, on the other hand, we  expect that there is no converging contour satisfying \eqref{geometric contour}.
It would be interesting to check $a)$ and $b)$ explicitly using  examples other than the figure-eight knot case in \eqref{figure-eight resurgence} and \eqref{figure-eight state-integration}.

In the below, we  give the explicit form of  unique converging cycle for several cases and numerically evaluate the central charges using the `short-cut' \eqref{Z^{conj} from state-integral}. For better reliability, it is recommended to do central charge computation again using the Borel resummation of the perturbative invariants which calls for  an efficient algorithm of computing the invariants, such as Feynman diagrams \cite{Dimofte:2012qj}.

\paragraph{Weeks manifold $=(S^3 \backslash {\bf 5^2_1})_{(5,-1)(5,-2)} = ({m003})_{(5,-2)}$ } {\it Weeks} manifold is the smallest volume hyperbolic 3-manifold. The state-integral is given by\footnote{We replace  the integration variables $(Z,X)$ in the state-integral model by $(bZ,bX)$ to make the symmetry $b\leftrightarrow b^{-1}$ manifest in the integrand.} (sloppy in the overall factor of the form \eqref{intrinsic ambiguity})
\begin{align}
&\CZ^{\overline{\textrm{hyp}}} \big{(}{\it Weeks};\hbar=2\pi ib^2 \big{)}   \nn
\\
&= \int_{\Gamma^{\overline{\textrm{hyp}}}_{{\it Weeks}}}\frac{dZ_1 dZ_2d X}{(2\pi)^3 \sqrt{2} } \big{(}2\cosh(\frac{bX}2 ) \cosh(\frac{X}{2b} ) \big{ )} \psi_b (Z_1 )\psi_b(Z_2) \nn
\\
& \quad \quad \times e^{-\frac{1}2 (b+b^{-1}) (2Z_1+Z_2+2X) - \frac{i}{4\pi}(4Z_1^2+Z_2^2 +4Z_2 X +3X^2+2Z_1 Z_2 + 8 Z_1 X) }  \;, \nn
\\
&\simeq \int_{{\gamma}^{\overline{\textrm{hyp}}}_{{\it Weeks}}}\frac{dZ_1 d X}{(2\pi)^2 \sqrt{2} } \big{(}2\cosh(\frac{bX}2 ) \cosh(\frac{X}{2b}) \big{ )} \psi_b (Z_1 )\psi_b(2X+Z_1) \nn
\\
& \qquad  \times \exp \big{[}- (b+b^{-1}) (Z_1+X) - \frac{i}{4\pi}(4Z_1^2 +3X^2 + 8 Z_1 X) \big{]} \;. \nn
\end{align}
Using an identity  of QDL  \eqref{FT of QDL}, we first integrated out  $Z_2$ along a cycle $\mathbb{E}_{2X+Z_1}$. The contour $\Gamma^{\overline{\textrm{hyp}}}_{\it Weeks} $ is a bundle over a 2d cycle $\gamma^{\overline{\textrm{hyp}}}_{\rm Weeks} \subset \mathbb{C}^2_{X,Z_1}$ whose fiber is the $\mathbb{E}_{2X+Z_1}$:
\begin{align}
\mathbb{E}_{2X+Z_1}\; \longrightarrow \;\;&\Gamma^{\overline{\textrm{hyp}}}_{Weeks} \nn
\\
& \downarrow \label{Bundle structure of contour}
\\
& \gamma^{\overline{\textrm{hyp}}}_{\it Weeks} \nn
\end{align}
One particular choice of  converging contour $\gamma^{\overline{\textrm{hyp}}}_{\it Weeks}$ in the reduced  two-dimensional integration is
\begin{align}
&\gamma^{\overline{\textrm{hyp}}}_{\it Weeks} := \big{\{} \big{(}Z_1, X \big{)}=\big{(}m_1 +i A_{\it Weeks}(m_1,m_2),m_2+i B_{\it Weeks}(m_1,m_2)\big{)}\; : \; (m_1,m_2) \in \mathbb{R}^2 \big{\}} \subset \mathbb{C}^2\;, \nn
\end{align}
where the continuous functions $A_{\it Weeks}(m_1, m_2)$ and $B_{\it Weeks}(m_1, m_2)$ have following asymptotic behavior :
\begin{align}
&\big{\{} A_{\it Weeks},B_{\it Weeks} \big{\}} =  \begin{cases} \big{\{}  \frac{(b+b^{-1})}8 ,0 \big{\}}  &\mbox{if } m_1 \geq \Lambda\;\;\; \;\;\textrm{   and }\;\;\; m_2 \geq \Lambda  \\ 
 \big{\{}  \frac{27(b+b^{-1})}{20},- \frac{11(b+b^{-1})}{20} \big{\}} &\mbox{if } m_1 \geq \Lambda \;\; \;\;\;\textrm{   and } \;\;\;m_2<-\Lambda \\ 
 \big{\{} \frac{(b+b^{-1})}8 , {\rm min}(\frac{|m_1|}{2 |m_2|},1) (b+b^{-1}) \big{\}} &\mbox{if } m_1 \leq -\Lambda\;\;\textrm{ and }\;\;\; m_2 \geq \Lambda \\ 
 \big{\{} 2(b+b^{-1}),-\frac{1}2 (b+b^{-1}) \big{\}} &\mbox{if } m_1 \leq -\Lambda \;\; \textrm{ and }\;\;\;m_2 \leq -\Lambda  \\ \end{cases} ,\nn
\end{align} 
with a proper positive number $\Lambda$, say $5$. For other asymptotic regions, the functions $(A_{Weeks},B_{Weeks})$ are  given by a linear interpolation of the above. For example, 
\begin{align}
&A_{\it Weeks}(m_1, m_2)=  \frac{1}{2\Lambda} (m_1 +\Lambda)A_{\it Weeks}(\Lambda, m_2)+ \frac{1}{2\Lambda}(\Lambda-m_1)A_{\it Weeks}(-\Lambda, m_2)   \;,  \nn
\\
&\textrm{when }-\Lambda\leq m_1\leq \Lambda  \textrm{ and } m_2 \geq \Lambda\;.
\end{align}  
The function can be continuously extended to the remaining finite region $[-\Lambda,\Lambda]^2 \subset \mathbb{R}^2$ without touching poles, see \eqref{poles of QDL}, in the integrand.
Since the integrand is locally holomorphic, small deformations of the contour do not change the final integration. The final result  only depends on an homology class of the contour and the extension to the finite region is unique as an element of the homology. Using the contour, we  numerically compute
\begin{align}
c_T (T[{\it Weeks}]) &=- \frac{8}{\pi^2} \textrm{Re} \bigg{[}\frac{\partial_b^2 \CZ^{\overline{\textrm{hyp}}} \big{(}{\it Weeks};\hbar=2\pi ib^2 \big{)} }{ \CZ^{\overline{\textrm{hyp}}} \big{(}{\it Weeks};\hbar=2\pi ib^2 \big{)}} \bigg{]}_{b=1}\;, \nn
\\
&\simeq 0.93\;.
\end{align}
\paragraph{Thurston manifold $=(S^3\backslash {\bf 5^2_1})_{(5,-1),(1,-2)} = (m003)_{(1,-2)}$} It is the second smallest hyperbolic closed 3-manifold. After integrating $Z_2$ using the identity  \eqref{FT of QDL}, the state-integral model reduced to
\begin{align}
&\CZ^{\overline{\textrm{hyp}}} \big{(}{\it Thurston};\hbar=2\pi ib^2 \big{)}  \nn
\\
&= \int_{{\gamma}^{\overline{\textrm{hyp}}}_{\it Thurston}}\frac{dZ_1 d X}{(2\pi)^2 \sqrt{2} } \big{(}2\cosh(\frac{bX}2 ) \cosh(\frac{X}{2b}) \big{ )} \psi_b (Z_1 )\psi_b(2X+Z_1) \nn
\\
& \qquad  \times \exp \big{[}- (b+b^{-1}) (Z_1+X) - \frac{i}{4\pi}(4Z_1^2 +7X^2 + 8 Z_1 X) \big{]}  \;.
\end{align}
The converging contour can be constructed in the same way as for $M={\it Weeks}$ case  using
\begin{align}
\big{\{} A_{\it Thurston},B_{\it Thurston} \big{\}} =  \begin{cases} \big{\{}  \frac{(b+b^{-1})}8 ,0 \big{\}}  &\mbox{if } m_1 >\Lambda\;\;\; \;\;\textrm{   and }\;\;\; m_2>\Lambda  \\ 
 \big{\{}  \frac{(b+b^{-1})}{2}, \frac{3(b+b^{-1})}{4} \big{\}} &\mbox{if } m_1 >\Lambda \;\; \;\;\;\textrm{   and } \;\;\;m_2<-\Lambda \\ 
 \big{\{} 2(b+b^{-1}) , - \frac{7}{8}(b+b^{-1}) \big{\}} &\mbox{if } m_1<-\Lambda\;\;\textrm{ and }\;\;\; m_2>\Lambda \\ 
 \big{\{} \frac{(b+b^{-1})}4 ,(b+b^{-1}) \big{\}} &\mbox{if } m_1 <-\Lambda \;\; \textrm{ and }\;\;\;m_2<-\Lambda  \\ \end{cases} . \nn
\end{align} 
Using the contour, we numerically obtain
\begin{align}
c_T (T[{\it Thurston}]) \simeq 1.01\;.
\end{align}
\paragraph{(5,-1)-Dehn filling on $m003$, $(S^3\backslash {\bf 5^2_1})_{(5,-1),(5,-1)} = (m003)_{(5,-1)}$ } The   reduced state-integral model for this case is
\begin{align}
&\CZ^{\overline{\textrm{hyp}}} \big{(}m003_{-5} ;\hbar=2\pi ib^2 \big{)}  \nn
\\
&= \int_{{\gamma}^{\overline{\textrm{hyp}}}_{m003_{-5}}}\frac{dZ_1 d X}{(2\pi)^2  } \big{(}2\sin(bX ) \sin(bX/2) \big{ )} \psi_b (Z_1 )\psi_b(2X+Z_1) \nn
\\
& \qquad  \times \exp \big{[}- (b+b^{-1}) (Z_1+X) - \frac{i}{4\pi}\big{(}4Z_1^2 -2X^2 + 8 Z_1 X \big{)} \big{]}  \;. \nn
\end{align}
For the contour, we use 
\begin{align}
\big{\{} A_{m003_{-5}},B_{m003_{-5}} \big{\}} =  \begin{cases} \big{\{}  \frac{(b+b^{-1})}8 ,0 \big{\}}  &\mbox{if } m_1 >\Lambda\;\;\; \;\;\textrm{   and }\;\;\; m_2>\Lambda  \\ 
 \big{\{}  2(b+b^{-1}), -\frac{9(b+b^{-1})}{10} \big{\}} &\mbox{if } m_1 >\Lambda \;\; \;\;\;\textrm{   and } \;\;\;m_2<-\Lambda \\ 
 \big{\{} \frac{(b+b^{-1})}8 ,(b+b^{-1}) \big{\}} &\mbox{if } m_1<-\Lambda\;\;\textrm{ and }\;\;\; m_2>\Lambda \\ 
 \big{\{}2(b+b^{-1}) ,-\frac{(b+b^{-1})}2 \big{\}} &\mbox{if } m_1 <-\Lambda \;\; \textrm{ and }\;\;\;m_2<-\Lambda  \\ \end{cases} . \nn
\end{align} 
Using the contour, numerically we find
\begin{align}
c_T (T[(m003)_{(5,-1)}]) \simeq 1.28\;.
\end{align}

\paragraph{Integral Dehn fillings on $m003$, $(S^3\backslash {\bf 5^2_1})_{(5,-1),(p,1)} = (m003)_{(p,1)}$ with $p \geq 5$ } The   reduced state-integral model is
\begin{align}
&\CZ^{\overline{\textrm{hyp}}} \big{(}m003_{p} ;\hbar=2\pi ib^2 \big{)}  \nn
\\
&= \int_{{\gamma}^{\overline{\textrm{hyp}}}_{m003_{p}}}\frac{dZ_1 d X}{(2\pi)^2  } \big{(}2\sinh(bX ) \sin(X/b) \big{ )} \psi_b (Z_1 )\psi_b(2X+Z_1) \nn
\\
& \qquad  \times \exp \big{[}- (b+b^{-1}) (Z_1+X) - \frac{i}{4\pi}\big{(}4Z_1^2 +(8+2p)X^2 + 8 Z_1 X \big{)} \big{]}  \;. \nn
\end{align}
One particular choice of   $\gamma^{\overline{\textrm{hyp}}}_{m003_p}$ ($p\geq 5$) is
\begin{align}
&\gamma^{\overline{\textrm{hyp}}}_{m003_p} := \bigg{\{} \big{\{}Z_1, X \big{\}}=\big{\{}m_1 +(b+b^{-1}) i ,m_2- \frac{2(b+b^{-1})i }{3\pi} \arctan(m_2)\big{\}}\; : \; m_1,m_2 \in \mathbb{R} \bigg{\}} \subset \mathbb{C}^2\;. \nn
\end{align} 
\\

\subsection{Comments on BPS operator spectrum} \label{sec : BPS spectrum}
 Here we present some difficulties in determining chiral operator spectrum of wrapped M5-brane theory $T[M]$. 
 The difficulties are closely related to the  two  challenges in 3d/3d story posed  in \cite{Chung:2014qpa}: a) recovering full flat connections on $M$ some of which are missing in ideal triangulations b) categorification/refinement of complex Chern-Simons theory.
 
\paragraph{Superconformal index computation} Superconformal index (SCI) is a physical quantity which contains information about BPS operator spectrum. Via the 3d/3d relation, it is also related to a complex CS ptn \cite{Dimofte:2011py,Gang:2013sqa,Yagi:2013fda,Lee:2013ida} as $S^3_b$-ptn. But there is a crucial difference between two  CS theories corresponding to  SCI and $S^3_b$-ptn: whether the non-quantized CS level $\sigma$ is real or purely imaginary,
\begin{align}
&\sigma \in i \mathbb{R}\; \quad \textrm{for superconformal index}\;, \nn
\\
&\sigma \in  \mathbb{R}\; \quad \;\; \textrm{for $S^3_b$-ptn with real $b$} \;. \nn
\end{align}
For {\it purely imaginary} $\sigma$, unitarity structure of the complex CS theory  is usual and $\bar{\CA}$ con be considered as  the complex conjugation of $\CA$. In the case, {\it all flat connections contribute} to the CS ptn  since they are on the contour, $\bar{\CA}= \CA^*$ \cite{Witten:2010cx} (see also the sec 2.1 of \cite{Gang:2015wya}).
 However, the construction based on ideal triangulation misses some branches of flat-connections including  Abelian branch \cite{Chung:2014qpa}. Thus, the state-integral based on ideal triangulation can not  give correct full SCI of $T[M]$.  {\it For real $\sigma$}, on the other hand, the situation is more subtle and it is {\it difficult to say which flat connections  may contribute} to the CS ptn from  view-point of purely complex CS theory.  Our basic assumption motivated by physical principles (holography and unitarity of wrapped M5-branes SCFTs) is that  the $\CA^{\overline{\textrm{hyp}}}$  is the only relevant flat connection appearing in the 3d/3d relation for $S^3_b$-ptn, see  eq.~\eqref{contour for real sigma} and \eqref{3d/3d in resurgence}. The flat connection is always captured in an ideal triangulation and the $S^3_b$-ptn can be computed using  the `incomplete' state-integral as we proposed in the previous sections. 
 
\paragraph{Non-trivial mixing of the IR R-charge}\footnote{We thank K. Yonekura, M. Yamazaki and N. Kim for pointing out this issue and subsequent discussions.} One may think the superonformal IR $R$-symmetry $U(1)^{\rm IR}_{R}$ of  3d SCFT $T[M]$ comes from a $SO(2)$ subgroup of the $SO(5)$ R-symmetry in the original 6d $(2,0)$ theory. If it is the case,  spectrum of the  $U(1)^{\rm IR}_R$ should be quantized and so should $\Delta_\Phi$ (conformal dimension of a chiral primary $\Phi$). Actually, however, the correct 3d IR $R$-charge  is a linear combination of $U(1)_R$ and $U(1)_t$ of $T^{\rm full}[M]$, see Table \ref{Comparison between two wrapped M5 SCFTs}. 
\begin{align}
\bigg{(} U(1)^{\rm IR}_{R} \textrm{ of }T[M] \bigg{)} \;\subset\; \bigg{(} U(1)_R \times U(1)_t  \textrm{ of }T^{\rm full}[M] \bigg{)}\;.
\end{align}
The correct IR mixing can be determined after identifying the  `Higgsing/deformation'  procedure in eq.~\eqref{T^{full}[M] to T[M]}. 
The complex mass parameter for the $U(1)_t$ plays a role as a refinement parameter in  refined  Chern-Simons theory \cite{Chung:2014qpa}.
Holographically the mixing can be studied by analyzing KK spectrum  on the holographic dual background \eqref{gravity dual of wrapped M5s}.  There is an unique massless vector field on the $AdS_4$ background which is given by a linear combination of the form \cite{Donos:2010ax} 
\begin{align}
A^{\rm IR} = A + \frac{3}{g}C\;,
\end{align}
where $g$ is a  constant in the supergravity solution related to the number of M5-branes. $A$ is an one-form obtained by  KK-reduction of   metric  along the $S^1$ isometry direction in the  $S^4$ and $C$ is an one-form  by KK-reduction of four-form field strength $G_4 =dC_3$  along a 3-cycle inside the internal manifold $M\times S^4$. The holographic dual of  $T^{\rm full}_N[M]$ should have two massless vector fields, $A$ and $C$,  corresponding to $U(1)_R$ and  $U(1)_t$ respectively. Since  an integration of  the Ramond-Ramond 4-form flux $G_4 = dC_3$ measures  M2-brane charge, the $U(1)_t$-charge counts M2-branes wrapping a 3-cycle inside $M\times S^4$.
It is compatible with the interpretation of $U(1)_t$ in  \cite{Gukov:2016gkn}.

\section{Kink CFTs in superconformal bootstrap} \label{sec: conformal bootstrap}

In this section, we analyze the SCFTs constructed in the previous section by wrapped two M5-branes on closed hyperbolic 3-manifolds in the eye of numerical superconformal bootstrap.
Studies on numerical bootstrap for 3d $\CN=2$ SCFT was given in~\cite{Bobev:2015vsa,Bobev:2015jxa}.
One of the notable features in their work was existence of three kinks for $\bar \Phi \Phi$ operator dimensions bounds.
From the numerical bootstrap studies of Ising model and $O(N)$ vector model, there are good indications that interesting theory (known or unknown) exists with CFT spectrum at the kink location~\cite{ElShowk:2012ht,El-Showk:2014dwa,El-Showk:2013nia,Kos:2013tga,Kos:2014bka,Kos:2015mba,Kos:2016ysd}.
Sudden change of numerical bounds indicate rearrangement of operator spectrum, such as certain operator decoupling, indicating interesting physical theory associated with it~\cite{El-Showk:2014dwa}.

The identity of the first kink is well described by critical Wess-Zumino(cWZ) model at $\Delta_\Phi = \frac{2}{3}$ whereas the second kink at $\Delta_\Phi =\frac{3}{4}$ is suggested to be coming from kinematic constraints~\footnote{Although, the fact that the second kink disappears for $d<3$ even though kinematic constraint still remains argue for something special about the second kink. The final word has not been set yet.}. 
The third kink seems interesting as it appears to show similar features as other interesting theories identifiable in the study of conformal bootstrap.
As far as we know,  there are currently no good identification to any known SCFT construction for the third kink. 
Similar exotic kink also appears in 4d $\CN=1$ SCFT~\cite{Poland:2011ey,Poland:2015mta}, and attempts to construct candidate theory for the kink~\cite{Xie:2016hny, Buican:2016hnq}.

First note that, we can easily exclude the possibility that the first kink SCFT (cWZ) as an wrapped M5-branes CFT $T[M]$.
Comparing the $S^3_b$ free-energy at small $b$:
\begin{align}
&\lim_{b\rightarrow 0}2\pi b^2 \CF_b (\textrm{cWZ}) = \textrm{Re}[i \Li_2 (e^{-2i \pi /3})]\simeq 0.6766 \nn
\\
&<\textrm{vol}(weeks) = 0.9427  \leq \textrm{vol}(M) = \lim_{b\rightarrow 0}2\pi b^2 \CF_b (T[M]) \textrm{ for any $M$} \nn
\\
&\Rightarrow \lim_{b\rightarrow 0}2\pi b^2 \CF_b (\textrm{cWZ model}) < \lim_{b\rightarrow 0}2\pi b^2 \CF_b (T[M]) \textrm{ for any $M$}
\end{align}
In the second line, we use  eq.~\eqref{two evidences} and the fact that the {\it Weeks} manifold has smallest hyperbolic volume.
We note that this argument does not exclude the possibility that the minimal SCFT can be realized   as a SCFT of generalized wrapped M5-branes system, such as including `irregular' co-dimension two defects in 6d $(2,0)$ theory along whole 3d spacetime and a link $K$ inside a closed 3-manifold $M$.

Now we focus on possibility whether SCFTs constructed by wrapped M5-branes on hyperbolic 3-manifolds are candidates for the SCFT associated with third kink .

\subsection{Third kink SCFT in superconformal boostrap}
Let us first review, the salient properties of the third kink observed in~\cite{Bobev:2015vsa,Bobev:2015jxa}:

\begin{itemize}
\item Chiral primary operator $\Phi$ with dimension $\Delta_\Phi \approx 0.86$.
\item Central charge of the kink solution is $ c_T \approx 0.93$.
\item  $\Phi$ has a chiral ring relation $\Phi^2 \sim 0$.
\end{itemize}
Recall that normalization for $c_T$ is given as in \eqref{normalization of central charge}.

In obtaining the numerical bounds authors of~\cite{Bobev:2015vsa,Bobev:2015jxa} performed moderate numerics in order to observe global patterns in large parameter space with various spacetime dimensions.
Here moderate numerics means in the sense of space of linear functional $\alpha$ in running numerical bootstrap\footnote{Note~\cite{Bobev:2015vsa,Bobev:2015jxa} use slightly different parameterization of linear functional, however we observe that their numerics are similar to $\Lambda =13$ in our notation, both in results and number of independent components.}
 (for a review see~\cite{Rychkov:2016iqz, Simmons-Duffin:2016gjk,Poland:2016chs})\begin{equation}
\alpha(F) = \sum_{m+n \leq \Lambda} a_{mn} \partial^m_z \partial^n_{\bar z} F(z, \bar z) \Big|_{z=\bar z = \tfrac{1}{2}}
\end{equation}
where $\Lambda$ is a cutoff introduced to make the problem finite. We are interested in constraints for $\Lambda \rightarrow \infty$. If numerics converges fast moderate value of $\Lambda$ can be sufficient, however there are cases(e.g.~\cite{Beem:2014zpa, Beem:2015aoa,Poland:2015mta,Lin:2015wcg,Collier:2016cls,Lin:2016gcl})  where reasonable computation do not yield converging result and requires extrapolation. 

We focus on studying the numerical bounds close to the third kink and report more stringent bounds.
Both  upper bound on $\Delta_{\bar \Phi \Phi}$  and  lower bound on $c_T$ vary as derivative order $\Lambda$ increases.
The strategy is to obtain bounds at multiple high $\Lambda$ and extrapolate to infer the value of bounds if we searched for infinite space of linear functionals.

For running the numerics obtained in this section, we used the semi-definite programming formulation of the problem~\cite{Poland:2011ey, Kos:2013tga}, making use of the solver  {\tt{SDPB}}~\cite{Simmons-Duffin:2015qma} and convenient wrapper {\tt{cboot}}~\cite{cbootOhtsky}.

\begin{figure} 
\begin{center} 
\includegraphics[height=2.5in]{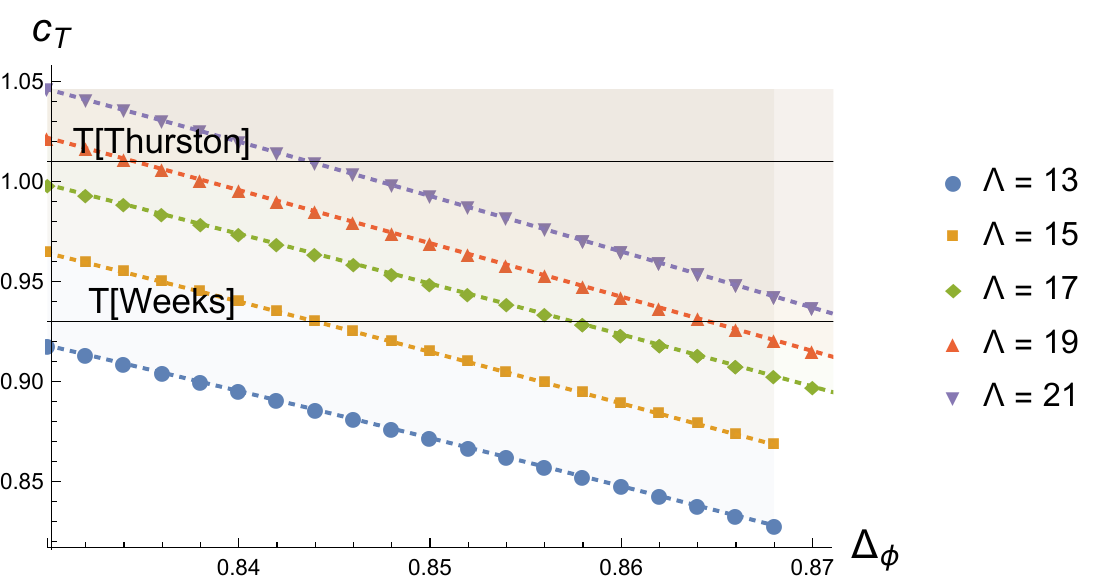}
\end{center}
\caption{Lower bound of $c_T$ for 3d $\CN=2$ SCFTs near the third kink point. The bound is obtained with most general assumption about SCFT spectrum consistent with unitarity. There is no interesting feature even at the kink point. For reference we included $c_T$ values computed for the $T[{\it Weeks}]$ and  $T[{\it Thurston}]$ wrapped M5-brane SCFTs obtained in the previous section.}
\label{cTNaiveBound}
\end{figure}

\subsection{Detailed analysis of the third kink CFT data}

In obtaining more stringent bound for $c_T$, we obtain bound directly on $c_T$ as opposed to extracting it from the extremal solution.
The naive $c_T$ bound from most generic unitary spectrum do not show interesting feature and monotonically decreases as $\Delta_\Phi$ passes through the third kink (see Figure~\ref{cTNaiveBound}). 
This is most generic bound on the central charge, however it fails to capture the information of extremal theories saturating the bound.
We can add extra assumption on the scalar spectrum that lightest $\Delta_{\bar \Phi \Phi}$ is maximal value obtained by the numerical bootstrap bound.
This is similar to studying the extremal spectrum and indeed we observe the same characteristics of the third kink with extremal spectrum study was done~\cite{Bobev:2015jxa} for $\Lambda = 13$.

\begin{figure}
\begin{subfigure}{.5\textwidth}
  \centering
  \includegraphics[height=1.6in]{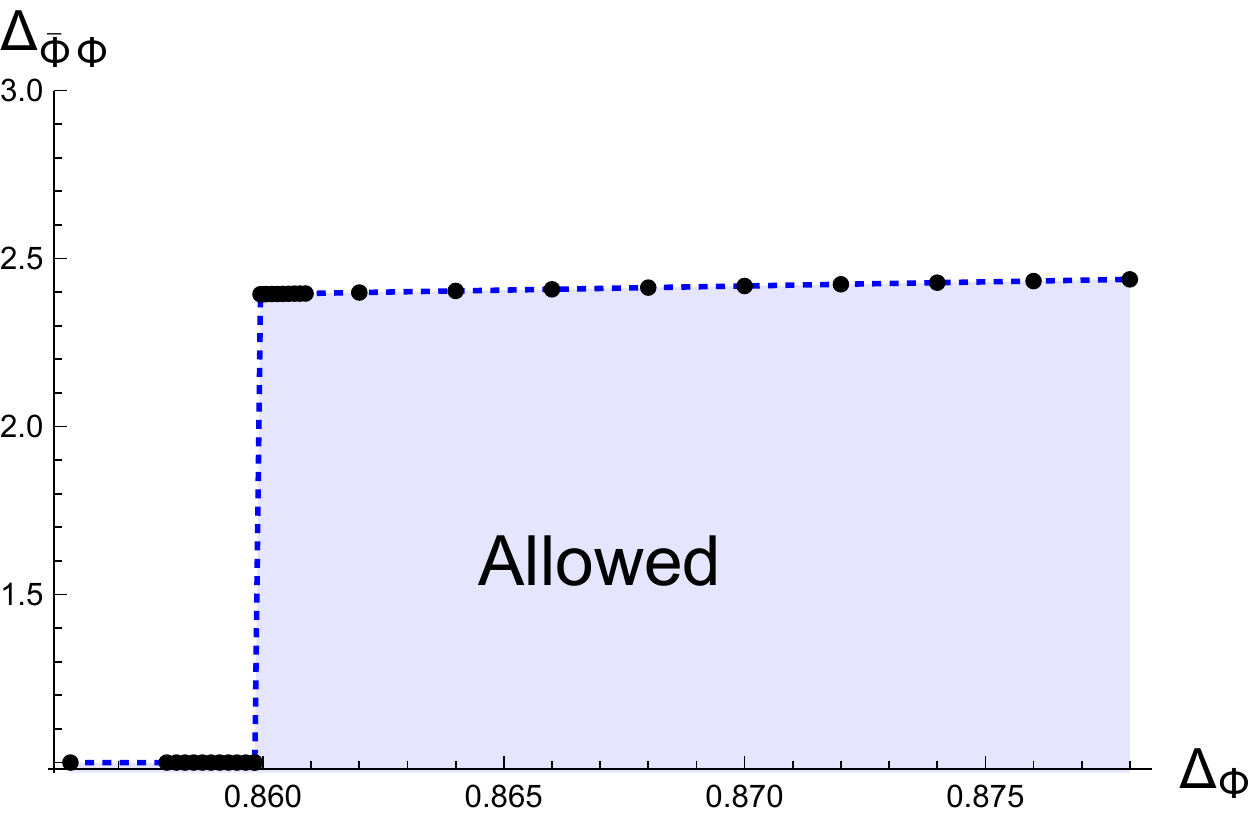}
  \caption{}
  \label{fig2:sfig1}
\end{subfigure}%
\begin{subfigure}{.5\textwidth}
  \centering
  \includegraphics[height=1.8in]{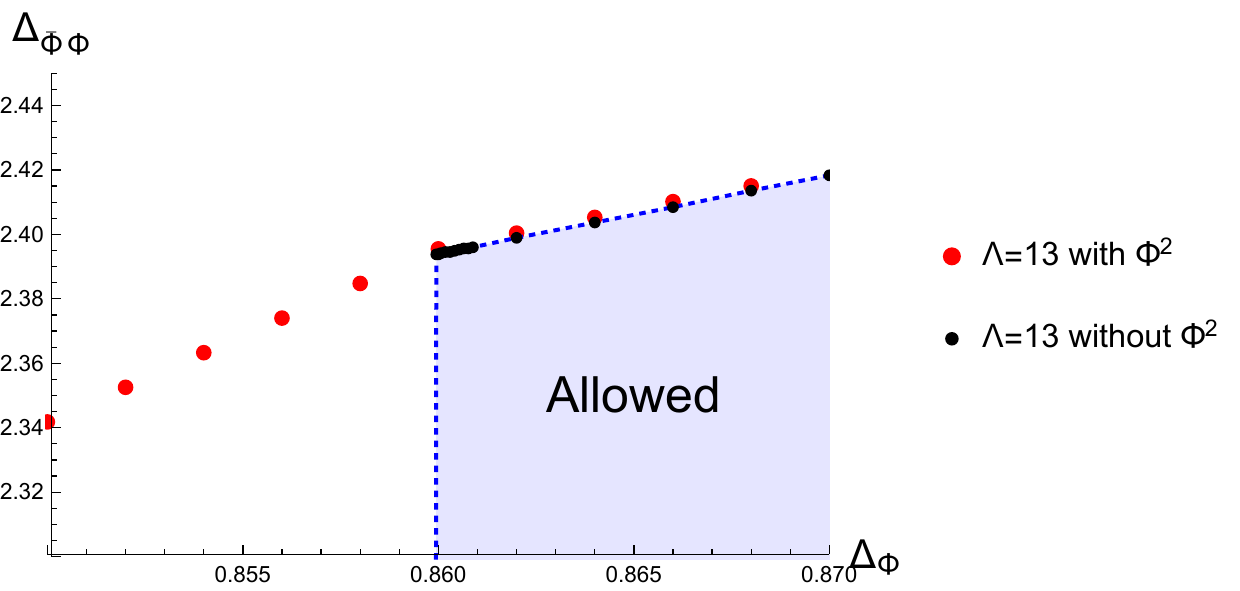}
  \caption{}
  \label{fig2:sfig2}
\end{subfigure}
\caption{(a) Upper bound for dimension of $\bar \Phi \Phi$ with explicit assumption of $\Phi^2$ operator decoupling. The jump is at
$(\Delta_\Phi, \Delta_{\bar \Phi \Phi}) = (0.8598(2),2.3937(5)).$ (b) Same bound overlapped with $\Phi^2$ included in the spectrum (red dots). The third kink can be identified as sudden jump (black dots) in the bound when spectrum excludes $\Phi^2$.  The numerics was obtained with $\Lambda =13$.}
\label{fig:fig}
\label{NoPhiJump}
\end{figure}

We used two different ways of identifying the third kink. First is the standard way of locating where the slope of the bound changes in $\Delta_{\bar \Phi \Phi}$ bound or $c_T$ bound with maximal $\Delta_{\bar \Phi \Phi}$ gap imposed. Another method is to use the fact that at the kink $\Phi^2$ chiral primary operator decouples. Extra assumption of excluding $\Phi^2$ operator in the SCFT spectrum drastically change the numerical bounds near the kink. This strategy was utilized in studying similar exotic kink for 4d $\CN=1$ SCFTs~\cite{Poland:2015mta}. This extra assumption allows us more efficient ways to study the spectrum of the kink as we can pin down the location with binary search in both $\Delta_\Phi$ and $\Delta_{\bar \Phi \Phi}$ directions. The reason that we could do binary search in $\Delta_\Phi$ is due to the bound having a jump at the kink (see Figure~\ref{NoPhiJump}). Near the jump point $\Delta_{\bar \Phi \Phi}$ close to unitarity bound is allowed if $\Delta_\Phi$ is greater than the kink location and disallowed if $\Delta_\Phi$ is smaller than the kink location.
We observe that the two approaches essentially gives the same result(e.g. Figure~\ref{NoPhiJump}) in identifying the kink, and therefore focus on the result from the second approach imposing $\Phi^2$ decoupling.

\begin{figure} 
\begin{center} 
\includegraphics[height=2in]{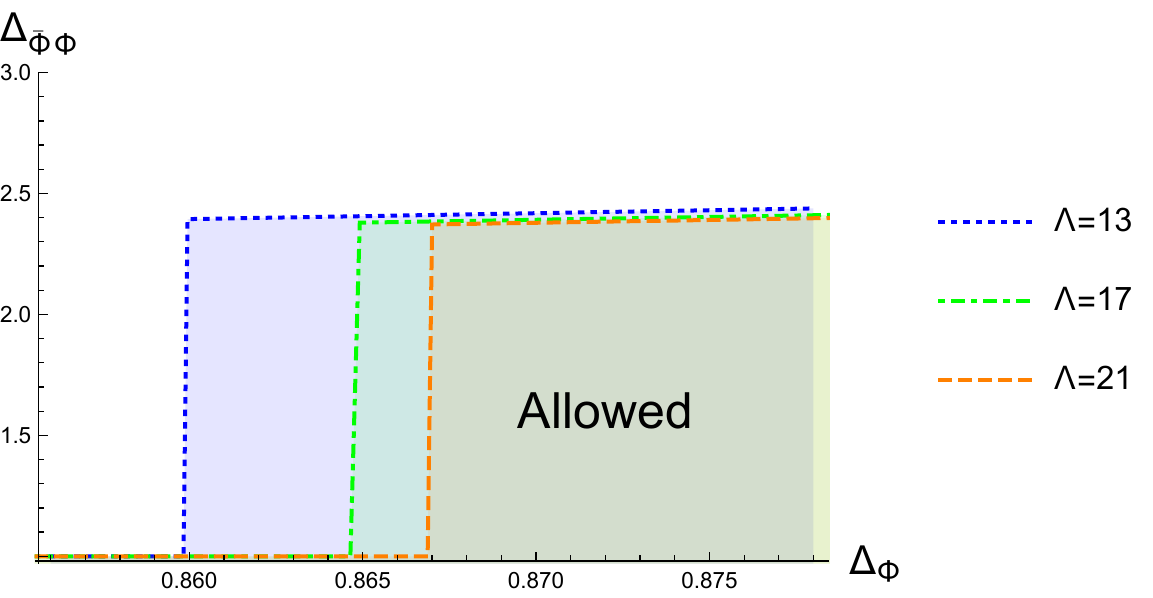}
\end{center}
\caption{Assuming no $\Phi^2$ in the spectrum, position of the third kink shifts to larger values as $\Lambda$ increases. The numerics does not converge at reasonable order and we resort to extrapolate to infer the location at $\Lambda \rightarrow \infty$. At a given order, the kink location can be efficiently searched with binary search over $\Delta_\Phi$ near where the jump occurs. }
\label{NoPhiJumpLambda}
\end{figure}

\begin{figure}
\begin{subfigure}{.33\textwidth}
  \centering
  \includegraphics[height=1.25in]{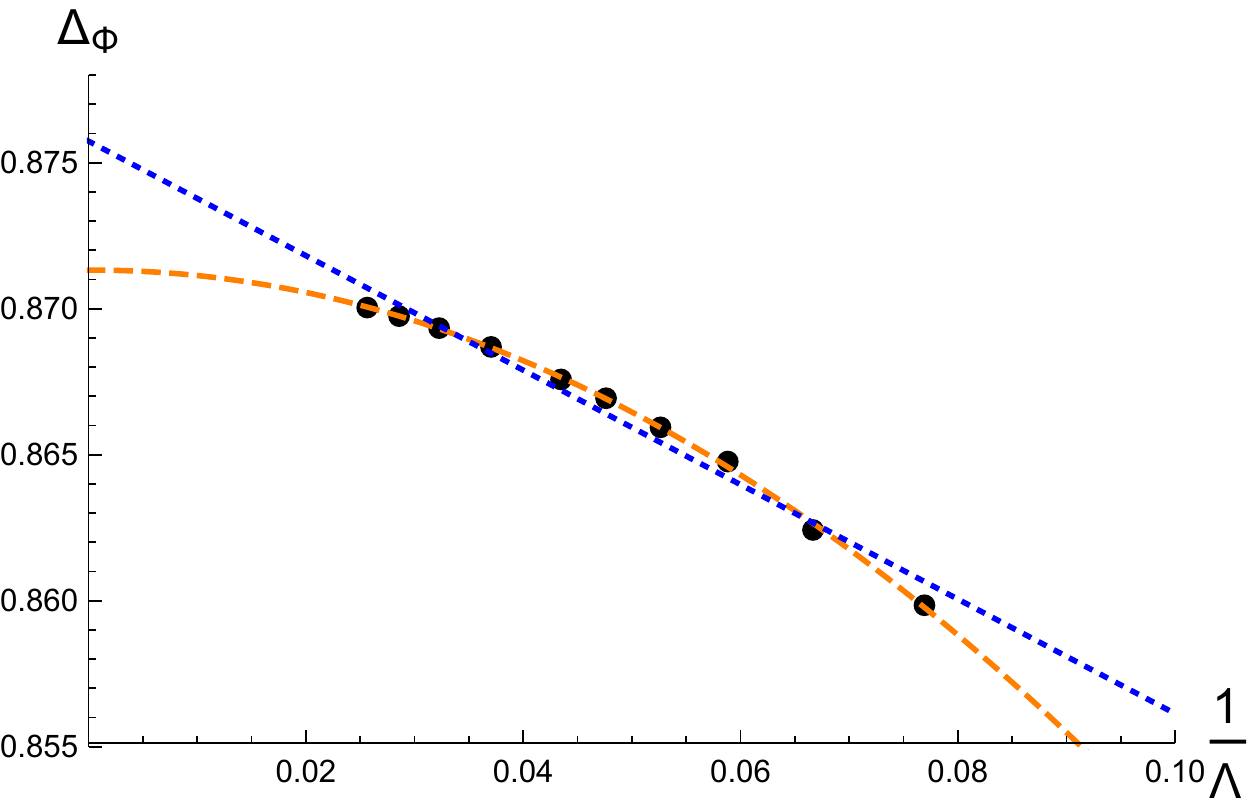}
  \caption{}
  \label{nophiextrapolate:multi:sfig1}
\end{subfigure}%
\begin{subfigure}{.33\textwidth}
  \centering
  \includegraphics[height=1.25in]{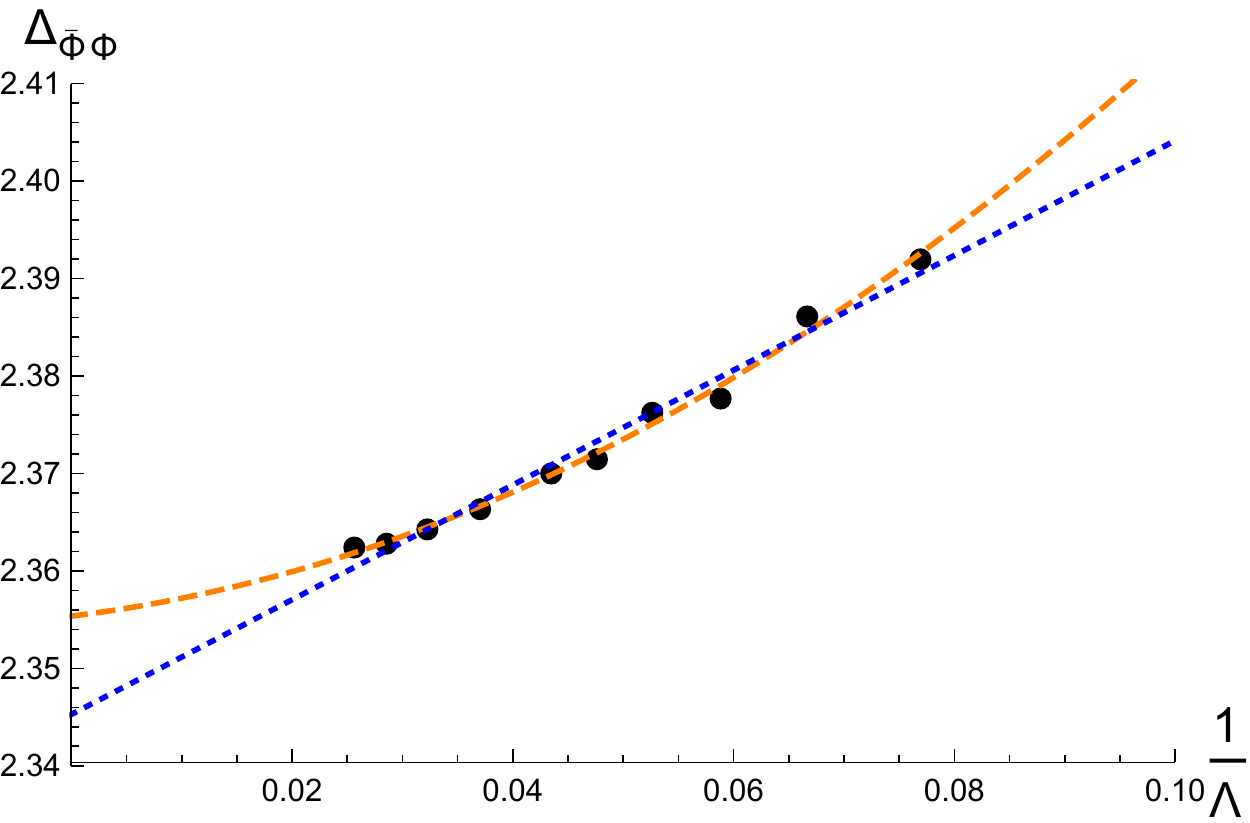}
  \caption{}
  \label{nophiextrapolate:multi:sfig2}
\end{subfigure}
\begin{subfigure}{.34\textwidth}
  \centering
  \includegraphics[height=1.25in]{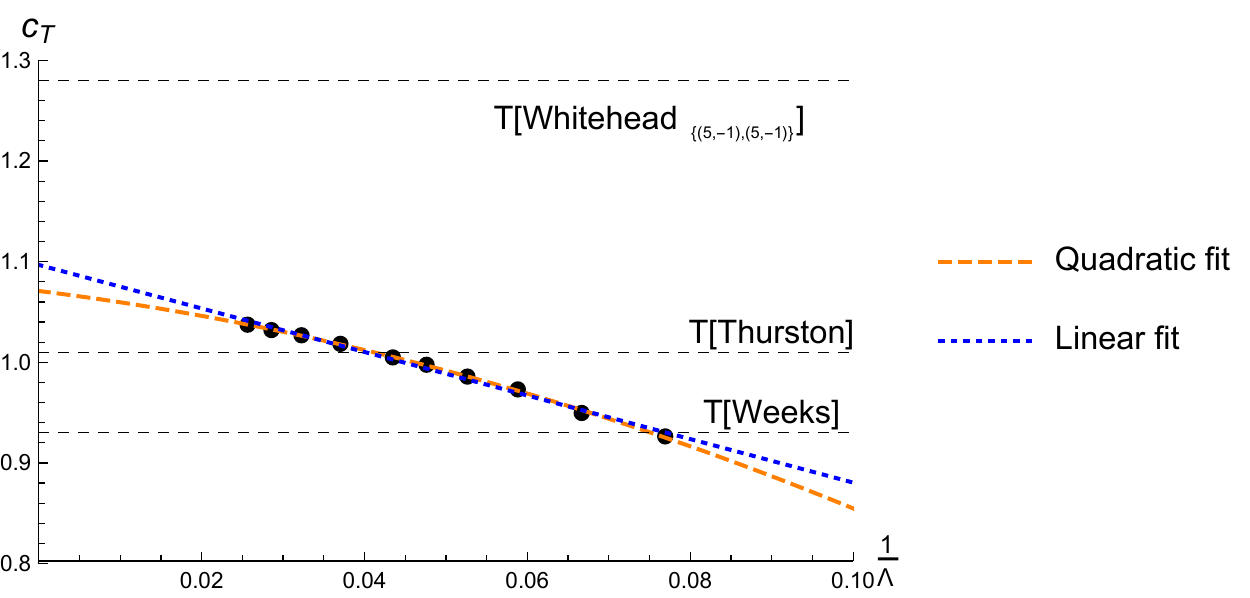}
  \caption{}
  \label{nophiextrapolate:multi:sfig3}
\end{subfigure}
\caption{CFT data at the third kink as function of $\frac{1}{\Lambda}$. (a) Lower bound of location of $\Delta_\Phi$ (b) Upper bound on $\Delta_{\bar \Phi \Phi}$ (c) Lower bound on central charge with maximal $\Delta_{\bar \Phi \Phi}$ gap(value in (b)) imposed.  The blue dashed-dotted line corresponds to the linear fit and orange dashed line corresponds to the quadratic fit of the points.  Refer to Table~\ref{table:nophiKink} for the asymptotic value at $\tfrac{1}{\Lambda} \rightarrow 0$. For reference we included $c_T$ values computed for three small hyperbolic volume wrapped M5-brane SCFTs obtained in the previous section. }
\label{nophiextrapolate:multi}
\end{figure}

First thing to observe is that the numerics does not converge as well as the first kink in the case for the third kink. For example, the location of the third kink shifts significantly as $\Lambda$ increases as see in Figure~\ref{NoPhiJumpLambda}. Other CFT data, which relies on the location of the kink, also varies as one increases derivative order. As the numerics do not converge at reasonable order, we extrapolate to infer the CFT data for $\Lambda \rightarrow \infty$. For the $c_T$ with maximal  $\Delta_{\bar \Phi \Phi}$ gap imposed (as well as other CFT data) see Figure~\ref{nophiextrapolate:multi} and Table~\ref{table:nophiKink}.

\renewcommand\arraystretch{1.5}
\begin{table}[h]
\centering
\begin{tabular}{ c c c c c } \hline\hline
\hspace{0.5cm} $\Lambda$ \hspace{0.5cm} &\hspace{0.5cm} $\Delta_\Phi$ \hspace{0.5cm} &\hspace{0.5cm} $c_T$(no gap) \hspace{0.3cm} &\hspace{0.3cm} $c_T$(maximal gap) \hspace{0.5cm} & \hspace{0.5cm} $\Delta_{\bar \Phi \Phi}$ \hspace{0.5cm} \\ \hline
13 & 0.8598(2) & $0.9258 $& 0.9264&  $2.3920(10)$  \\
15& 0.8624(2) & $0.9485$ & $0.9497$ & $2.3861(10)$ \\
17 & 0.8647(2) & $0.9717$ & $0.9731$ & $2.3777(10)$ \\
19 & 0.8659(2) & $0.9842$ & $0.9858$ & $2.3762(10)$  \\
21 & 0.8669(2) & $0.9962$ & $0.9976$ & $2.3715(10)$  \\
23 & 0.8676(2) & $1.0034$ & $1.0049$ & 2.3700(10)  \\
27 & 0.8687(2) & $1.0172$ & $1.0183$ &  2.3663(10)  \\
31 & 0.8693(2) & $1.0257$ & $1.0269$ & 2.3643(10) \\
35 & 0.8697(2) & $1.0318$ & 1.0320 & 2.3628(10)  \\
39 & 0.8700(2) & $1.0361$  & 1.0374 & 2.3624(10) \\ \hline\hline
$\infty$ & $0.8757/0.8713$ & $1.0957/1.0723$ & $1.0968/1.0709$ & $2.3453/2.3554$
\end{tabular}
\caption{Few CFT data of the third kink at various $\Lambda$ assuming no $\Phi^2$ in the spectrum. The bottom row indicates interpolated values for $\Lambda \rightarrow \infty$, which should be taken as suggested value rather than a strict bound. Interpolation was taken with both linear(left value) and quadratic(right value) fit respect to $\frac{1}{\Lambda}$. The values for given $\Lambda$ gives strict upper bound for $\Delta_{\bar \Phi \Phi}$ and strict lower bound for $\Delta_\Phi$ and $c_T$. }
\label{table:nophiKink}
\end{table}

According to our computation, both $T[{\it Weeks}]$  and $T[{\it Thurston}]$ are ruled out as the candidate for the third kink SCFT irrespective of interpolation(see Figure~\ref{nophiextrapolate:multi}).
Initial study from~\cite{Bobev:2015jxa} indicated a reasonable match with the $c_T$ value of $T[{\it Weeks}]$ SCFT. However, further analysis with higher order numerics  excludes the potential identification. 
It would be interesting to check if $T[M]$ with other 3-manifolds with bigger volume can be identified as the 3rd kink.

{\small
\section*{Acknowledgements}

We would like to thank M.~Yamazaki, K.~Yonekura, N.~Kim, T.~Ohtsuki, Y.~Nakayama, S.~Maeda,  C.M-Thompson,  M.~Romo, T. Dimofte,  J.~Cho, S.~Kim, S.~Pufu, S.~Rychkov and H.~Chung for invaluable discussion and encouragement.
The contents of this paper was presented by DG\footnote{Nagoya U (Dec 2015), SNU (Dec 2015, Aug 2016), KIAS (Dec 2015),  NTU (Jan 2016), IPMU (Feb 2016), PMI (Aug 2016) and Waseda U (``Workshop on
Volume Conjecture and Quantum Topology", Sep 2016). }, and we thank the audience for feedback.
The research of DG is supported in part by the WPI Initiative (MEXT, Japan).
The research of DG is supported by the JSPS-NRF collaboration program with grant No. NRF-2016K2A9A2A08003745. 
DG is also supported by a Grant-in-Aid for Scientific Research on Innovative Areas 2303 (MEXT, Japan). JB thanks to KIAS Center for Advanced Computation for providing computing resources.
}

\appendix

\section{Ideal triangulation of $S^3\backslash {\bf 5^2_1}$ and $m003$} \label{App : NZ-datum}
Ideal triangulations of 3-manifolds with cusped boundaries are available in a computer software  {\tt SnapPy} \cite{snappy}. 
\paragraph{Whitehead link complement ($S^3\backslash {\bf 5^2_1}$)} The 3-manifold can be triangulated by 4 ideal tetrahedrons ($T=4$). Boundary meridian/longitude variables and indepedent internal edges $(\mathcal{C})$  are
\begin{align}
& X_1 = -Z_1''-Z_3''+Z_4 \; \;, \;\; P_1 = \frac{Z_1}{2}-Z_3+\frac{3 Z_4}{2}-\frac{Z'_1}{2}-\frac{Z'_4}{2}-Z''_1-Z''_3-\frac{Z''_2}{2} \;,\nn
\\
& X_2 = Z_1- Z_2'-Z_3 \;\;, \;\; P_2 = Z_1-\frac{Z_3}{2}-\frac{Z_4}{2}+\frac{Z'_3}{2}-\frac{Z'_1}{2}-\frac{Z'_2}{2}-\frac{Z''_4}{2} \;,\nn
\\
&\mathcal{C}_1  = 2Z_1'+Z_1''+2Z_2'+Z_2''+Z_3+Z_4''-2\pi i  \;,\nn 
\\
&\mathcal{C}_2 = Z_1''+Z_2''+2Z_3''+Z_3+2Z_4'+Z_4''-2\pi i \;.
\end{align}
Using a linear relation
\begin{align}
Z_i +Z_i'+Z_i''=i \pi\;,
\end{align}
the edge parameter $Z'_i$ can be eliminated. 
After the elimination, generalized Neumann-Zagier datum $(A,B,C,D;f, f'',\nu,\nu_p)$ are determined by
\begin{align}
&A \cdot   \left(\begin{array}{c}Z_1 \\ Z_2 \\ Z_3 \\ Z_4  \end{array}\right) + B  \cdot \left(\begin{array}{c}Z''_1 \\ Z''_2 \\ Z''_3 \\ Z''_4  \end{array}\right) - i \pi \nu = \left(\begin{array}{c}X_1 \\X_2 \\ \mathcal{C}_1 \\ \mathcal{C}_2  \end{array}\right) \;,  \;\;C \cdot \left(\begin{array}{c}Z_1 \\ Z_2 \\ Z_3 \\ Z_4  \end{array}\right)   + D  \cdot \left(\begin{array}{c}Z''_1 \\ Z''_2 \\ Z''_3 \\ Z''_4  \end{array}\right) - i \pi \nu_p = \left(\begin{array}{c} P_1 \\ P_2 \\ \G_1 \\ \G_2   \end{array}\right) \nn
\\
& A\cdot f+B\cdot f''  = \nu\;, \quad C\cdot f+D\cdot f'' = \nu_p\;. \nn
\end{align}
Here $\{\Gamma_i\}_{i=1}^2$ are  some linear combinations of $\vec{Z}$ and $\vec{Z}''$ chosen to satisfy 
\begin{align}
\left(\begin{array}{cc}A & B \\ C & D\end{array}\right) \in Sp(8, \mathbb{Q})\;.
\end{align}
For example, we can choose
\begin{align}
&\G_1 = \frac{Z_1}{2}-\frac{Z_3}{4}-\frac{Z_4}{4}+\frac{Z''_1}{4}-\frac{Z''_2}{2}-\frac{Z''_3}{4}-\frac{Z''_4}{4}\;, \nn
\\
&\G_2 =\frac{Z_1}{2}+\frac{Z_4}{2}-\frac{5 Z_3}{8}+\frac{3 Z''_1}{8}+\frac{Z''_4}{8}-\frac{3 Z''_2}{8}\;.
\end{align}
The  final expression of the state-integral model is independent on the specific choice of $\Gamma_1$ and  $\Gamma_2$.

\paragraph{Sister of figure-eight knot complement = $(S^3\backslash 5^2_1)_{{\bf (5,-1)}} = ( m003)$} The 3-manifold can be triangulated by 2 ideal tetrahedrons ($T=2$). After eliminating $(Z_1', Z_2')$, we have
\begin{align}
&X = Z_1''+Z_1-2Z_2-3Z_2'' + i\pi \;, \quad  P = -Z_2-2Z_2''+i \pi \;. \nn
\\
&\mathcal{C} = Z_1''+2Z_1-Z_2-2Z_2'' \;, \quad \Gamma = Z_1''+Z_1 \nn
\end{align}
Generalized Neumann-Zagier datum $(A,B,C,D;f, f'',\nu,\nu_p)$ are determined by
\begin{align}
&A \cdot   \left(\begin{array}{c}Z_1 \\ Z_2   \end{array}\right) + B  \cdot \left(\begin{array}{c}Z''_1 \\ Z''_2   \end{array}\right) - i \pi \nu = \left(\begin{array}{c}X \\ \mathcal{C} \end{array}\right) \;,  \;\;C \cdot \left(\begin{array}{c}Z_1 \\ Z_2 \end{array}\right)   + D  \cdot \left(\begin{array}{c}Z''_1 \\ Z''_2  \end{array}\right) - i \pi \nu_p = \left(\begin{array}{c} P  \\ \G  \end{array}\right) \nn
\\
& A\cdot f+B\cdot f''  = \nu\;, \quad C\cdot f+D\cdot f'' = \nu_p\;. \nn
\end{align}
\section{Quantum dilogarithm} \label{App : QDL}
In this appendix we collect formulas for the noncompact quantum dilogarithm (QDL) function \cite{FaddeevKashaevQuantum}. 
The function function  $\Psi_b (Z)$ is defined by 
\begin{align}
\Psi_b (Z) := \begin{cases} \prod_{r=1}^\infty \frac{1-q^r e^{-Z}}{1-\tilde{q}^{-r+1}e^{-\tilde{Z}}}&\mbox{if } |q|<1 \\ 
  \prod_{r=1}^\infty \frac{1-\tilde{q}^r e^{-\tilde{Z}}}{1-q^{-r+1}e^{-Z}} &\mbox{if } |q|>1 \\ 
\end{cases}
\end{align}
with
\begin{align}
q:=e^{2\pi i b^2}\;, \quad \tilde{q}:=e^{2\pi i b^{-2}}\;, \quad \tilde{Z}:= \frac{1}{b^2}Z\;.
\end{align}
Integral representation:
\begin{align} 
\log\Psi_{b}(Z)=\int_{\mathbb{R}+i 0^+}  \frac{e^{ \frac{ i t Z }{\pi b} + t (b+b^{-1})}}{\sinh(b t)\sinh(b^{-1}t)} \frac{dt}{4t}\;, \quad \textrm{for $0<\textrm{Im}[Z]<2\pi (1+b^{2})$\;.}
\end{align}
Asymptotic  expansion when $\hbar =2\pi i b^2 \rightarrow 0$ :
\begin{align}
\log\Psi_{b}(Z) \xrightarrow{\;\;b^2\rightarrow 0^+\;\;}   \sum_{n=0}^{\infty} \frac{B_n \hbar^{n-1}}{n!} \textrm{Li}_{2-n}(e^{-Z})\;, \quad \textrm{for $0<\textrm{Im}[Z]<\pi$} \;. \label{asymptotic of QDL}
\end{align}
Here $B_n$ is the $n$-th Bernoulli number with $B_1=1/2$.
To have $b\leftrightarrow b^{-1}$ symmetry, we define 
\begin{align} 
\log \psi_b (x):= \log \Psi_{b } (b x)  \;.
\end{align}
At $b=1$, the QDL simplified as
\begin{align}
\log \psi_{b=1}(x)  = \frac{-(2\pi +i x) \log (1-e^{-x})+i \textrm{Li}_2 (e^{-x})}{2\pi} \;.
\end{align}
As $|x|\rightarrow \infty$,
\begin{align}
\log \psi_b (x) \quad \sim \quad &\frac{-x^2}{4\pi i }+ \frac{1}2 (b+b^{-1}) x &\textrm{for Re[x]$<$0} \;,  \nn
\\
\sim \quad  &  0 &\textrm{for Re[x]$>$0}\;.
\end{align}
Poles of the $\psi_b (Z)$ are located on 
\begin{align}
\mathbb{Z}_{\leq 0 } (2\pi i b)+\mathbb{Z}_{\leq 0} (2\pi i b^{-1})\; \;. \label{poles of QDL}
\end{align}
Fourier transformation: %
\begin{align}
&\frac{e^{ \frac{i\pi(b^2+3+b^{-2})}{12}}}{2\pi}\int_{\mathbb{E}_y} dx \psi_b (x) e^{\frac{x^2 + 2x y-2 \pi i x (b+b^{-1})}{4\pi i }} =  \psi_b(y) \quad \quad \textrm{for Im}(y)>0\;, \nn
\\
&\mathbb{E}_y := \{ x+i f(x)\;:\; x\in \mathbb{R}  \} \subset \mathbb{}\; \textrm{where $f$ is a function satisfying}\nn
\\
&f \rightarrow  \begin{cases} -\textrm{Im}(y)+(b+b^{-1})\pi -\epsilon_1 &\mbox{if } x \rightarrow  \infty \\ 
 \epsilon_2 &\mbox{if } x <\Lambda  \\ 
\end{cases} \textrm{ with small $\epsilon_1,\epsilon_2 >0$ and positive $\Lambda$}.\label{FT of QDL}
\end{align}

\section{Quantum Dehn filling}\label{App : Dehn's filling}
Classical phase space $P(\partial M)$ and its Lagrangian subvariety $\CL(M)$ for the $SL(2)$ CS theory  are
\begin{align}
&P(\partial M)  = \{ SL(2)\textrm{-flat connections on $\partial M = (\mathbb{T}^2)^{|K|-S}$} \} = \big{(}P(\mathbb{T}^2) \big{)}^{|K|-S}\;  \nn
\\
&  \;\; \qquad \qquad \textrm{with } P(\mathbb{T}^2) = (\mathbb{C}^*)^2/\mathbb{Z}_2 =\{(x,p)\in (\mathbb{C}^*)^2 : (x,p)\sim (1/x, 1/p) \} \;,\nn
\\
&\mathcal{L} (M)  = \{SL(2)\textrm{-flat connections on $ M$}  \} \;.\label{phase,Lagrangian}
\end{align}
Here $x$ and $p$ parametrize the $SL(2)$ gauge holonomy around each  meridian and longitude respectively:
\begin{align}
\textrm{P} e^{ \oint_{\textrm{merdian}} \CA} = \left(\begin{array}{cc} x & 1 \\0 & 1/x\end{array}\right)\;, \quad  \textrm{P} e^{\oint_{\textrm{longitude}} \CA} = \left(\begin{array}{cc} p & 1 \\0 & 1/p\end{array}\right)\;.
\end{align}
Quantizing them, we have
\begin{align}
P(\partial M)\; &\leadsto\; \CH(\partial M) = \big{(}\CH (\mathbb{T}^2)\big{)}^{|K|-S}\;\;  \textrm{(a Hilbert-space)}\;,  \nn
\\
 \CL( M)\;& \leadsto \; \big{|}\CZ(M) \big{\rangle} \in \CH(\partial M) \qquad \;\; \; \;\textrm{(a state)}\;.
\end{align}
%

\paragraph{Quantization of the phase space $P(\mathbb{T}^2)$ with $k=1$}
Phase space $\CP (\mathbb{T}^2)$ for $SL(2)_{k, \sigma}$ CS theory with $k=1$  and $\sigma= \frac{1-b^2}{1+b^2}$   on $\mathbb{R}_t \times \mathbb{T}^2$ is give in \eqref{phase,Lagrangian} with following symplectic form ($X:=\log x, P:=\log p$):
\begin{align}
 &\Omega = \frac{1}{\pi (1+b^2)}dP \wedge dX + \frac{1}{\pi (1+b^{-2})} d \overline{P}\wedge d \overline{X} \;.
 \end{align}
Quantization of the  phase space give an infinite dimensional Hilbert-space $\CH (\mathbb{T}^2)$ whose position basis are 
\begin{align}
\textrm{Position bais of $\CH(\mathbb{T}^2)$} = \big{\{} |X\rangle \;:\; X\in \mathbb{C}\;, \;|X\rangle \sim = |-X\rangle\big{\}}.
\end{align}
The quantum position/momentum operators acts on the Hilbert-space as %
\begin{align}
\langle X| \hat{x}= \langle X| e^{X}  \;, \quad \langle X| \hat{\bar{x}}= \langle X| e^{X/b^2}  \;, \quad \langle X| \hat{p}= \langle X+i \pi b^2| \;, \quad  \langle X| \hat{\bar{p}}= \langle X+i \pi | \;.
\end{align}
Completeness relation in $\CH (\mathbb{T}^2)$  is 
\begin{align}
& \frac{1}{4 \pi b}  \int d \mu  \big{|}X \big{\rangle} \big{ \langle} X \big{|} = \mathbb{I}  \;.\label{completeness relation}
\end{align}
\paragraph{Quantization of Dehn filling}

For a 3-manifold closed $M$ obtained by gluing two 3-manifolds $M_1$ and $M_2$ along a common  $\mathbb{T}^2$  boundary   with a $\varphi \in SL(2,\mathbb{Z})$ twist, the $SL(2)$ CS ptn is given by 
\begin{align}
&\CZ_{\rm SI} (M = M_1 \cup_{\varphi} M_2;\hbar) = \big{\langle} \CZ(M_1) \big{|} \hat{\varphi}  \big{|}\CZ(M_2 )\big{ \rangle}\;,  \nn
\\
& \big{|}\CZ(M_i)  \big{\rangle} \in \CH(\mathbb{T}^2)\;, \quad i=1,2\;, \nn
\\
&\hat{\varphi} \; : \; \CH(\mathbb{T}^2) \rightarrow \CH(\mathbb{T}^2)\;.
\end{align}
For solid-torus $D_2\times S^1$, the wave-function is simply given by %
\begin{align}
\big{\langle} X \big{|} \CZ(D_2\times S^1) \big{\rangle} = 4  \sinh(X )\sinh(X/b^2) \;.
\end{align}
Note that solid-torus can be thought as unknot complement on $S^3$, $D^2\times S^1 =S^3\backslash \mathbf{0}_1$, and we use the canonical polarization where the position (momentum) is an eigenvalue homonomy around the meridian (longitude).
The wave-function satisfy a pair of difference equations ($q:=e^{2\pi i b^2}, \bar{q}:=e^{2\pi i b^{-2}}$):
\begin{align}
&\hat{A}_{K=\mathbf{0}_1} (\hat{x}^2,\hat{p}, q^{1/2}) \big{|} \CZ(D^2\times S^1)\big{\rangle}= \hat{A}_{K=\mathbf{0}_1} (\hat{\bar{x}}^2,\hat{\bar{p}}, \bar{q}^{1/2}) \big{|} \CZ(D^2\times S^1)\big{\rangle} =0\;, \nn
\\
&
\textrm{where }\hat{A}_{\mathbf{0}_1} (\hat{x}^2, \hat{p},q^{1/2}) =\hat{p}^2+1-q^{1/2}\hat{p}-q^{-1/2} \hat{p}\;. 
\end{align}
Regardless of whether the gauge group is $SU(2)$ or its complexification $SL(2)$, the difference operator $\hat{A}_K$ annihilating the knot-complement wave-function $|Z(S^3\backslash K)\rangle$ is the same and called `quantum A-polynomial' of knot $K$ \cite{Gukov:2003na}. For a closed 3-manifold $(S^3\backslash K)_{p/q}$ obtained by performing Dehn surgery with a slope $p/q$ on $S^3$ along a knot $K$\footnote{We call a link $K$ with one component ($|K|=1$) a  `knot'.}, the CS wave function  can be obtained as follows:
\begin{align}
&(S^3\backslash K)_{p/q} = (D^2\times S^1) \cup_{\varphi_{p/q}} (S^3\backslash K) \;, \quad  \varphi_{p/q}:=\left(\begin{array}{cc}* & * \\ p & q\end{array}\right)  \in SL(2,\mathbb{Z})\;,\nn
\\
& \CZ_{\rm SI} \big{(} (S^3\backslash K)_{p/q};\hbar\big{)} = \big{ \langle} \CZ(D^2\times S^1) \big{|} \hat{\varphi}_{p/q}\big{|} \CZ(S^3 \backslash K) \big{\rangle} \;, \; \quad \hat{\varphi}_{p/q}\; : \; \CH (\mathbb{T}^2) \rightarrow \CH (\mathbb{T}^2)\;.
\end{align}
Two generators of $SL(2,\mathbb{Z})$ are
\begin{align}
\varphi_S= \left(\begin{array}{cc}0 & -1 \\ 1 & 0\end{array}\right)\;, \quad \varphi_T = \left(\begin{array}{cc}1 & 0 \\ 1 & 1\end{array}\right)\;.
\end{align}
Quantization of these operators give \cite{Dimofte:2014zga} 
\begin{align}
&\hat{\varphi}_S \;, \; \hat{\varphi}_T\; :\; \CH(\mathbb{T}^2) \rightarrow  \CH(\mathbb{T}^2)\;, \nn
\\
& \big{\langle} X\big{|} \hat{\varphi}_S   \big{|} \psi \big{\rangle}  = \frac{1}{\sqrt{2} \pi b} \int dY e^{- \frac{ XY} {\pi i b^2} } \big{\langle} Y \big{|} \psi \big{\rangle} \;,  \nn
\\
&  \big{\langle} X \big{|} \hat{\varphi}_T  \big{ |}\psi \big{\rangle} =  e^{\frac{1 }{2 \pi i b^2} X^2} \big{\langle} X \big{|}\psi \big{\rangle}\;, \quad \textrm{for any $\big{|}\psi  \big{\rangle} \in \CH(\mathbb{T}^2)$}\;.
\end{align}
For general element $\varphi=\left(\begin{array}{cc}r & s \\ p & q\end{array}\right) \in SL(2,\mathbb{Z})$,
\begin{align}
&\big{\langle} X \big{|} \hat{\varphi} \big{|}\psi\rangle=   \frac{1}{\sqrt{2s} \pi b} \int dY e^{\frac{  q X^2}{2 \pi i  b^2 s}+\frac{  X Y }{\pi i b^2 s} +\frac{ r Y^2}{2 \pi i b^2 s}} \big{\langle} Y \big{|} \psi \big{\rangle}\;, \quad \textrm{for }s\neq 0\;,\nn
\\
& \big{\langle} X \big{|} \hat{\varphi}  \big{ |}\psi \big{\rangle} =  e^{\frac{ p X^2}{2 \pi i b^2 r} } \big{\langle} X \big{|}\psi \big{\rangle}
\;, \quad \textrm{for }s=0\;.
\end{align}
Inserting the completeness relation \eqref{completeness relation}, we have 
\begin{align}
& \CZ_{\rm SI} \big{(} (S^3\backslash K)_{p/q};\hbar\big{)}  =\big{ \langle} \CZ(D^2\times S^1) \big{|} \hat{\varphi}\big{|} \CZ(S^3 \backslash K) \big{\rangle}\nn
\\
& = \frac{1}{4\pi b} \int dX   \big{\langle}  \CZ(D^2\times S^1) \big{|} X \big{\rangle}  \big{\langle} X \big{|} \hat{\varphi}\big{|} \CZ(S^3\backslash K)\big{\rangle} \nn
\\
&= \frac{1}{ \pi^2 b^2\sqrt{2s}} \int dX dY \sinh(X )\sinh(X/b^2) e^{\frac{  q X^2}{2 \pi i  b^2 s}+\frac{  X Y }{\pi i b^2 s} +\frac{ r Y^2}{2 \pi i b^2 s}}   \big{\langle} Y \big{|} \CZ(S^3\backslash K)\big{\rangle} \nn
\\
&=\int \frac{\Delta_b(Y;s,q) dY}{(2 \pi q \hbar )^{1/2}}  \exp \big{(}\frac{p} { \hbar q}  Y^2\big{)} \CZ_{\rm SI}( S^3\backslash K;Y;\hbar)\;.
\end{align}
Here $\Delta_b$ is defined in eq.~\eqref{Dehns-filling}. For given $(p,q)$, the $s$ is  determined modulo $q\mathbb{Z}$ and the final expression $\CZ_{\rm SI} \big{(} (S^3\backslash K)_{p/q}\big{)}$ does not depend on the choice of $(r,s)$ modulo the intrinsic ambiguity \eqref{intrinsic ambiguity}. This is compatible with the fact that the resulting 3-manifold does not depends on $(r,s)$ but only on $(p,q)$.

\newpage

\bibliographystyle{JHEP}
\bibliography{w3c}

\end{document}